\documentclass[12pt,preprint]{aastex}

\newcommand{\beq}{\begin{equation}}
\newcommand{\eeq}{\end{equation}}
\newcommand{\beqa}{\begin{eqnarray}}
\newcommand{\eeqa}{\end{eqnarray}}
\newcommand{\bit}{\begin{itemize}}
\newcommand{\eit}{\end{itemize}}

\begin{document}

\title{Resonant Scattering and Ly-alpha Radiation Emergent from Neutral Hydrogen Halos}

\author{Ishani Roy\altaffilmark{1}, Chi-Wang Shu\altaffilmark{1}, and Li-Zhi
Fang\altaffilmark{2}}

\altaffiltext{1}{Division of Applied Mathematics, Brown University,
Providence, RI 02912} \altaffiltext{2}{Department of Physics,
University of Arizona, Tucson, AZ 85721}

\begin{abstract}

With a state-of-the-art numerical method for solving the
integral-differential equation of radiative transfer, we investigate
the flux of the Ly$\alpha$ photon $\nu_0$ emergent from an optically thick
halo containing a central light source. Our focus is on the
time-dependent effects of the resonant scattering. We first show
that the frequency distribution of photons in the halo are quickly
approaching to a locally thermalized state around the resonant
frequency, even when the mean intensity of the radiation is highly
time-dependent. Since initial conditions are forgotten during the
thermalization, some features of the flux, such as the two peak
structure of its profile, actually are independent of the intrinsic
width and time behavior of the central source, if the emergent
photons are mainly from photons in the thermalized state. In this
case, the difference $|\nu_{\pm}-\nu_0|$, where $\nu_{\pm}$ are the
frequencies of the two peaks of the flux, cannot be less than $2$
times of Doppler broadening. We then study the radiative transfer in
the case where the light emitted from the central source is a flash.
We calculate the light curves of the flux from the halo. It shows
that the flux is still a flash. The time duration of the flash for
the flux, however, is independent of the original time duration of
the light source but depends on the optical depth of the halo.
Therefore, the spatial transfer of resonant photons is a diffusion
process, even though it is not a purely Brownian diffusion. This
property enables an optically thick halo to trap and store
thermalized photons around $\nu_0$ for a long time after the cease
of the central source emission. The photons trapped in the halo can
yield delayed emission, of which the profile also shows typical two
peak structure as that from locally thermalized photons.  Possible
applications of these results are addressed.

\end{abstract}


\newpage

\section{Introduction}

The transfer of Ly$\alpha$ radiation in optically thick medium is
fundamentally important for the understanding of the physics of
halos around Ly$\alpha$ photon sources or clouds nearby these
sources. It includes Ly$\alpha$ forest, damped Ly$\alpha$ system,
Ly$\alpha$ blob, Ly$\alpha$ emitter and fluorescent Ly$\alpha$
emission of galaxies and quasars, as well as the optical afterglow
of gamma ray bursts. The profiles of the emission and absorption of
the Ly$\alpha$ radiation from these sources are powerful tools to
constrain the mass density, velocity, temperature and the fraction
of neutral hydrogen of IGM at various redshifts (Miralda-Escude \&
Rees 1998; Miralda-Escude 1998; Haiman \& Cen 2005; Tasitsiomi
2006; Totani et al. 2006; McQuinn et al. 2007).

It is well known that the resonant scattering of Ly$\alpha$ photons
with neutral hydrogen atoms has a profound effect on the time,
space and frequency dependencies of the transfer of Ly$\alpha$
photons. An analytical solution of the integro-differential
equation of the resonant radiative transfer revealed that the
resonant scattering leads to a local Boltzmann distribution of
photons in a small frequency range around the Ly$\alpha$ frequency
$\nu_0$ (Wouthuysen 1952; Field 1958, 1959). The temperature of the
Boltzmann distribution is equal to the kinetic temperature of the
neutral hydrogen atoms. The width of the local Boltzmann
distribution is increasing with time. This is the so-called
Wouthuysen-Field effect, which is important in $21$ cm cosmology (Fang
2009).

Besides Field's solution, no other time-dependent analytical
solutions of the integro-differential equation are available. All
other analytical solutions (Harrington 1973; Neufeld 1990; Dijkstra
et al. 2006) are based on the time-independent Fokker-Planck
equation, which is the diffusion approximation of the radiation
transfer. These time-independent solutions are important, but can
only be used to describe the ``limiting asymptotic behavior" of the
radiative transfer (Adams 1972). It gives nothing of the
time-development of the local Boltzmann distribution. The Monte
Carlo numerical method has also been used to solve the radiative
transfer of Ly$\alpha$ photons (Lee 1974, Zheng \& Miralda-Escude
2002; Ahn 2002; Cantalupo et al. 2005; Verhamme et al. 2006). Yet, the
time-development of the Wouthuysen-Field effect is also absent in
this approach.

Time-independent approximation would be reasonable if the length
scale $l$ and time scales $t$ of the problem considered satisfy the
condition $l/t \ll c$. In this case, one can take $c\rightarrow
\infty$, and then, the time derivative term $\partial/c\partial t$
of the radiative transfer equation can be dropped. The condition
$l/t \ll c$, however, cannot be satisfied if the size of the neutral
hydrogen halo is large and the time scale of the Ly$\alpha$ photon
source is small. For instance, the time scale $t_{\rm after}$ of the
optical afterglow of GRBs is of the order of a few tens of hours
(e.g. Tanvir et al. 2009; Salvaterra et al. 2009), or a few days
(Vreeswijk et al. 2004), while the column number density of neutral
hydrogen is as large as 10$^{21-22}$ cm$^{-2}$, and number volume
density is about 10$^2$ -10$^4$ cm$^{-3}$ (Vreeswijk et al. 2004).
Therefore, the size of the neutral hydrogen halos should be much
larger than $ct_{\rm after}$. It is improper to treat the short-time
problems with the solutions of ``limiting asymptotic behavior".

Recently, a state-of-the-art numerical solver for the kinetic
equations has been developed.  This solver is based on the weighted
essentially non-oscillatory (WENO) scheme (Jiang \& Shu 1996). It
has been developed to solve the Boltzmann equations (Carrillo et al.
2003, 2006) and radiative transfer (Qiu et al. 2006, 2007, 2008).
This numerical solver has successfully passed the tests of
conservation of photon number, Field's solutions and the
Wouthuysen-Field effect etc., and has been properly used as the
solver of the transfer of resonant photons (Roy et al. 2009a, b, c).

We will study, in this paper, the time-dependent behavior of
the Ly$\alpha$ radiation transfer in an optically thick medium. We will not
try to explain any specific observed Ly$\alpha$ spectrum, instead we will
study the physical features of the resonant photon transfer, which
can not be addressed with previous solvers. For instance, we will
show that the frequency distribution of the Ly$\alpha$ photons can keep
in a locally thermalized state even when the intensity and flux are
highly time-dependent. This feature is essentially important, as
thermalization generally will lead to the initial conditions being forgotten.
Our solver is also able to study the transfer of a light flash in optically
thick halo. It shows that the nature of the transfer is a diffusion process,
though it is not a purely Brownian diffusion. This property leads to the
trap and store of photons thermalized around the Ly$\alpha$ frequency
for a long time after the cease of the central source emission.

This paper is organized as follows. Section 2 presents the
theoretical background of the Ly$\alpha$ photon transfer in
an optically thick medium. Section 3 gives the solution of Ly$\alpha$
photons emergent from an optically thick spherical halos with a
steady source located at the center of the halo. Section 4
is on the solutions when the central source is a flash. A discussion
and conclusion are given in Section 5. The details of the numerical
implementation is given in the Appendix.

\section{Theory of Ly-alpha radiative transfers in optically thick halos}

\subsection{Optically thick halos}

The property of the halo around individual luminous object depends
on the luminosity, the spectrum of UV photon emission, and the
time-evolution of the center object. For luminous objects at high
redshift, the halos generally consist of three spheres (Cen 2006;
Liu et al. 2007). The most inner region is the highly ionized
Str\"omgren sphere, or the HII region, which is optically thin of
Ly$\alpha$ photons. The second region, which is just outside the HII
region, is optically thick of Ly$\alpha$ photons. The temperature of
the baryon gas is about $10^{4}$ K, which is due to the heating of the UV
photons. The third region is outside of the heated region. It is
un-heated, and therefore, the temperature of the baryon gas can be as
low as $10^2$ K.

In this context, we will study the transfer of Ly$\alpha$ photons in
a radius $R$ spherical halo of neutral hydrogen with temperature $T$
in the range of $10^2$ to $10^4$ K. Assuming  the uniformly
distributed HI gas has number density $n_{\rm HI}$, the optical
depth over a light path $dl$ is $d\tau(\nu)=\sigma(\nu)n_{\rm
HI}dl$, where $\sigma(\nu)$ is the cross section of the resonant
scattering of Ly$\alpha$ photons by hydrogen, given as
\begin{equation}
\sigma(x)=\sigma_0  \phi(x,a)
\end{equation}
where $x$ is the dimensionless frequency defined by $x\equiv
(\nu-\nu_0)/\Delta \nu_D$, $\nu_0$ being the resonant frequency.
$\Delta \nu_D=\nu_0 (v_T/c)$ is the Doppler broadening, and
$v_T=\sqrt{2k_BT/m}$. Therefore, $x$ measures the frequency
deviation $\Delta \nu=|\nu-\nu_0|$ in units of $\Delta \nu_D=
1.06\times 10^{11}(T/10^4)^{1/2}$ Hz. $\sigma_0=\pi e^2 f/m_e c
\Delta \nu_D =1.10 \times 10^{-2}$ cm$^{2}$ is the cross section of
the resonant scattering at the frequency $\nu_0=2.46\times10^{15}$
s$^{-1}$. The function $\phi(x,a)$ in equation (1) is the normalized
profile given by the Voigt function as (Hummer 1965)
\begin{equation}
\phi(x,a)=\frac{a}{\pi^{3/2}}\int^{\infty}_{-\infty} dy \frac
{e^{-y^2}}{(x-y)^2+a^2}.
\end{equation}
The parameter $a$ in equation (2) is the ratio of the natural to the
Doppler broadening. For the Ly$\alpha$ line, $a=4.7\times
10^{-4}(T/10^4)^{-1/2}$. The optical depth of the halo with column
number density of neutral hydrogen $N_{\rm HI}=n_{\rm HI}R$ is
\begin{equation}
\tau(x) = N_{\rm HI} \sigma(x) = \tau_0\phi(x,a).
\end{equation}
Since $\phi(0,a)=1/\sqrt{\pi}$ when $a\ll 1$, the line-center
optical depth is then $\tau(0)=\tau_0/\sqrt{\pi}$, and
\begin{equation}
\tau_0 = 1.04\times10^{7}\left (\frac{T}{10^4}\right )^{-1/2} \left
(\frac{N_{\rm HI}}{10^{20}{\rm cm^{2}}}\right).
\end{equation}

\subsection{Radiative transfer equation in spherical halo}

The radiative transfer of Ly$\alpha$ photons in spherical halo is
described by the equation of the specific intensity
$I(\eta,r,x,\mu)$ as
\begin{eqnarray}
\lefteqn{ {\partial I\over\partial \eta} + \mu \frac{\partial I}
{\partial r}+\frac{(1-\mu^2)}{r}\frac{\partial I}{\partial \mu}
- \gamma \frac{\partial I}{\partial x} = } \nonumber \\
 & &   - \phi(x;a)I + \int \mathcal{R}(x,x';a)I(\eta, r,
x',\mu')dx'd\mu' + S,
\end{eqnarray}
where we use dimensionless time $\eta$ defined as $\eta=cn_{\rm
HI}\sigma_0 t$ and dimensionless coordinate $r$ defined as $r=n_{\rm
HI}\sigma_0 r_p$, with $r_p$ being the physical radial coordinate.
That is,  $\eta$ and $r$ are, respectively, in the units of mean
free flight-time and mean free path of photon $\nu_0$. For a signal
propagated in the radial direction with the speed of light, we have
$r= \eta+ {\rm const}$. In equation (5) $\mu=\cos \theta$ is the
direction relative to the radial vector ${\bf r}$.

The re-distribution function $\mathcal{R}(x,x';a)$ gives the
probability of a photon absorbed at the frequency $x'$, and
re-emitted at the frequency $x$. It depends on the details of the
scattering (Henyey 1941; Hummer 1962; Hummer 1969). If we consider
coherent scattering without recoil, the re-distribution function
with the Voigt profile equation (2) is
\begin{eqnarray}
\lefteqn{ \mathcal{R}(x,x';a)= } \\ \nonumber
 & \ \ \  & \frac{1}{\pi^{3/2}}\int^{\infty}_{|x-x'|/2}e^{-u^2}
\left [
\tan^{-1}\left(\frac{x_{\min}+u}{a}\right)-\tan^{-1}\left(\frac{x_{\max}-u}{a}\right
)\right ]du
\end{eqnarray}
where $x_{\min}=\min(x, x')$ and $x_{\max}=\max(x,x')$. In the case
of $a=0$, i.e. considering only the Doppler broadening, the
re-distribution function is
\begin{equation}
\mathcal{R}(x,x')=\frac{1}{2}{\rm erfc}[{\rm max}(|x|,|x'|)].
\end{equation}
The re-distribution function of equation (7) is normalized as
$\int_{-\infty}^{\infty} \mathcal{R}(x,x')dx'=\phi(x,0)
=\pi^{-1/2}e^{-x^2}$. With this normalization, the total number of
photons is conserved in the evolution described by equation (5). That is,
the destruction processes of Ly$\alpha$ photons, such as the
two-photon process (Spitzer \& Greenstein 1951; Osterbrock 1962),
is ignored in equation (5). In equations (6) and (7), we also do not consider the
recoil of atoms. It is equal to assume the mass of atom is very
large. The effect of recoil actually is under control (Roy et al.
2009c). We will address it in next section.

In equation (5), the term with the parameter $\gamma$ is due to the expansion
of the universe. If $n_{\rm H}$ is equal to the mean of the number
density of cosmic hydrogen, we have $\gamma=\tau_{GP}^{-1}$, and
$\tau_{GP}$ is the Gunn-Peterson optical depth. Since Gunn-Peterson
optical depth is of the order of $10^{6}$ at high redshift (e.g. Roy
et al. 2009c), we will take the parameter $\gamma=10^{-5}-10^{-6}$.

\subsection{Eddington approximation}

When the optical depth is large, we can take the Eddington approximation
as
\begin{equation}
I(\eta,r,x,\mu)\simeq J(\eta, r, x) + 3\mu F(\eta, r,x)
\end{equation}
where $J(\eta,r,x)=\frac{1}{2}\int_{-1}^{+1}I(\eta,r,x,\mu)d\mu$ is
the angularly averaged specific intensity and
$F(\eta,r,x)=\frac{1}{2}\int_{-1}^{+1}\mu I(\eta,r,x,\mu)d\mu$ is
the flux. Defining $j=r^2J$ and $f=r^2F$, Eq.(5)  yields
the equations of $j$ and $f$ as
\begin{eqnarray}
{\partial j\over\partial \eta} + \frac{\partial f} {\partial r} & =
& - \phi(x;a)j + \int \mathcal{R}(x,x';a)j dx'+ \gamma
\frac{\partial
j}{\partial x}+ r^2S,\\
 \frac{\partial f}{\partial \eta} + \frac{1}{3}
\frac{\partial j} {\partial r} - \frac{2}{3}\frac{j}{r} & = & -
\phi(x;a)f.
\end{eqnarray}
The mean intensity $j(\eta,r,x)$ describes the $x$ photons
trapped in the halo by the resonant scattering, while the flux
$f(\eta,r,x)$ describes the photons in transit.

The source term $S$ in the equations (9) and (10) can be described by a
boundary condition of $j$ and $f$ at $r=r_0$. We can take $r_0=0$, as
$r_0$ is not important if the optical depth of the halo is large. Thus,
we have
\begin{equation}
j(\eta, 0, x)=0, \hspace{1cm} f(\eta, 0, x)=S_0\phi_s(x),
\end{equation}
where $S_0$, and $\phi_s(x)$ are, respectively, the intensity and
normalized  frequency profile of the sources. Since equation
(9)-(11) are linear, the intensity $S_0$ can be taken as any constant. That
is, the solution $f(x)$ of $S_0=S$ is equal to $S f_1(x)$, where $ f_1(x)$ is the solution
of $S=1$. On the other hand, the equations (9) and (10) are not linear with
respect to the function $\phi_s(x)$, i.e. the solution $f(x)$ of $\phi_s(x)$
is not equal to $\phi_s(x)f_1(x)$, where $f_1$ is the solution of $\phi_s(x)=1$.

In the range outside the halo, $r>R$, no photons propagate in the direction
$\mu<0$. Therefore, the boundary
condition at $r=R$ given by $\int_{0}^{-1}\mu J(\eta, R, x, \mu)d\mu
=0$ is then (Unno 1955)
\begin{equation}
j(\eta, R,x)=2f(\eta, R,x).
\end{equation}
There is no photon in the field before $t=0$. Therefore, the initial
condition is
\begin{equation}
j(0,r,x)=f(0,r,x)=0.
\end{equation}

We solve equations (9) and (10) with the numerical method developed
recently (Roy et al. 2009a, 2009b, 2009c). Some details of this
method is given in the Appendix. We first solve the problems when
the sources $S$ is steady.

\section{Solutions of steady sources}

\subsection{Time scale of escape}

First we consider steady sources. That is, the parameter $S_0$ in
eq.(11) is time-independent after the switch-on of the sources at
$\eta=0$ [eq.(13)]. A typical solution of the flux $f(\eta, r,x)$
given by equations (9) and (10) is shown in Figure 1, for which the
source is taken to be $S_0=1$ and
$\phi_s(x)=(1/\sqrt{\pi})e^{-x^2}$, i.e. the emission line width is
equal to the Doppler broadening. The parameters $a$ and $\gamma$ are
taken to be $10^{-3}$ and $10^{-5}$, respectively. The effect of
$\gamma=10^{-5}$ actually is ignorable in these solutions. The left
panel is the solutions $f(\eta, r, x)$ at radius $r=10^2$ and time
$\eta=500$, $1000$, $2000$ and 3000 with the boundary condition
equations (11) and (12). The solutions approach to a stable state at
time $\eta \geq 2000$.

\begin{figure}[htb]
\begin{center}
\includegraphics[scale=0.28]{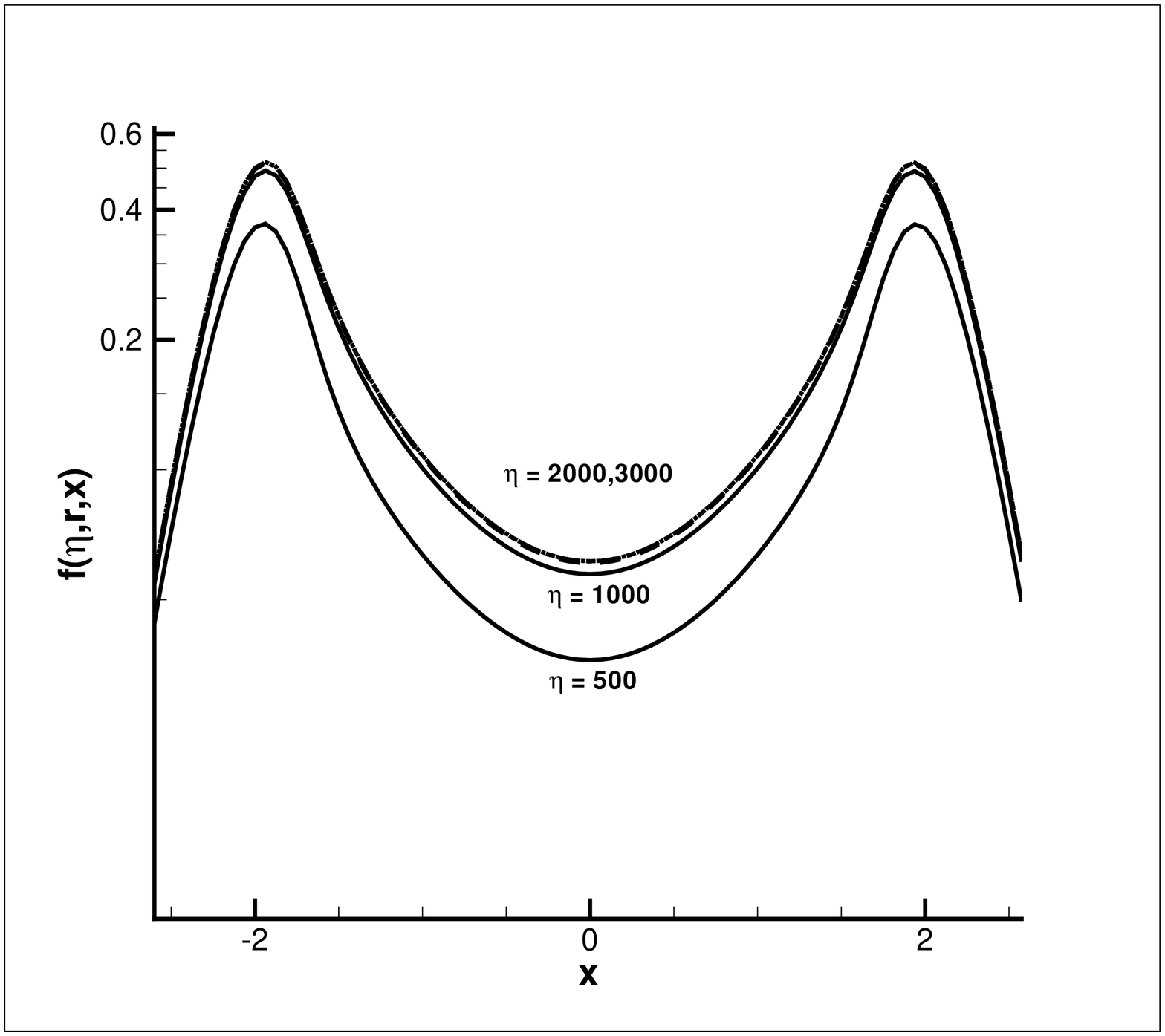}
\includegraphics[scale=0.28]{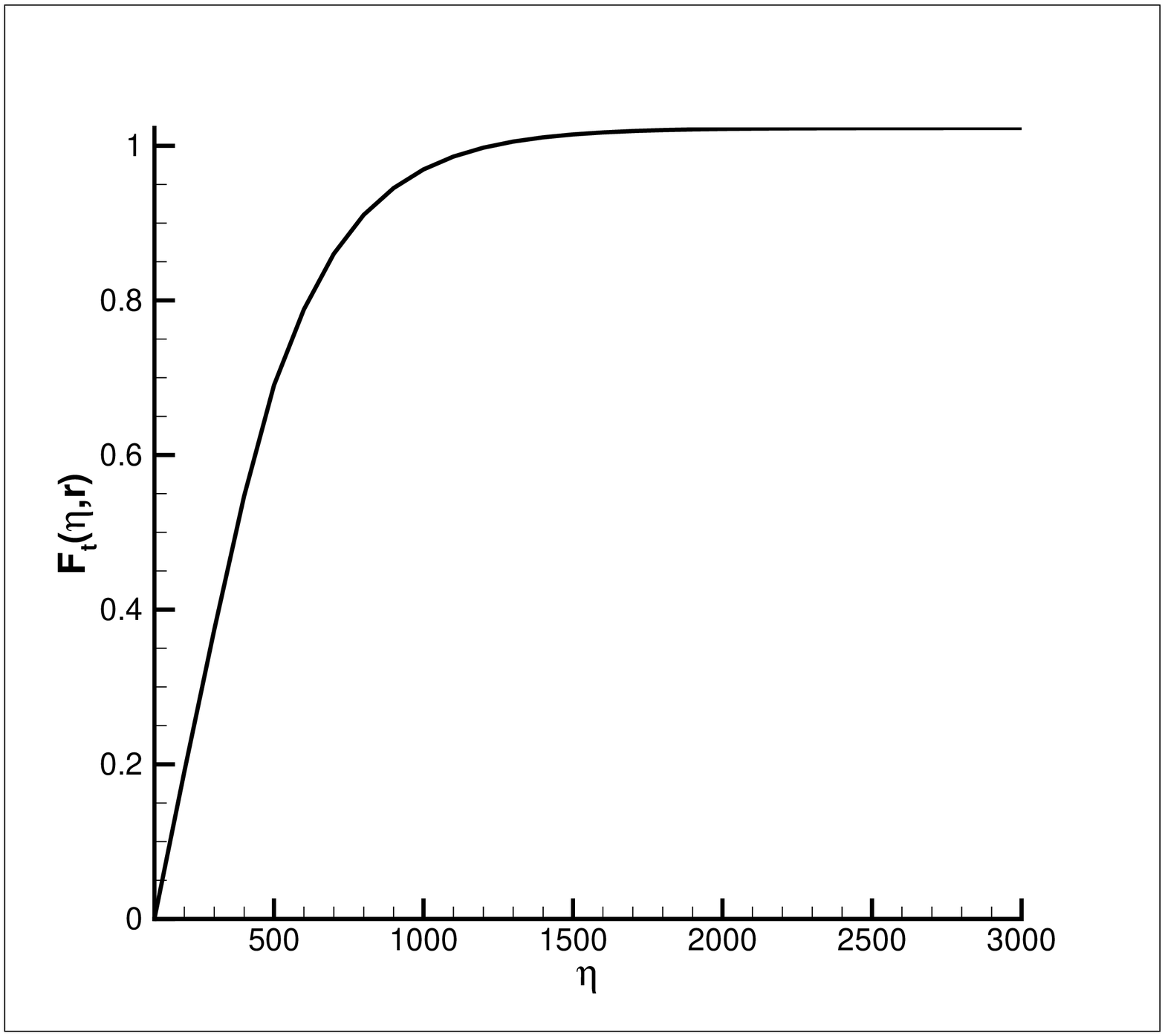}
\end{center}
\caption{Left panel is the solution of flux $f(\eta, r, x)$ of
equations (9) and (10) at $r=R=10^2$ and $\eta=500$, $1000$, $2000$
and 3000 with the boundary condition equation (12), $R=10^2$. The
source is $S_0=1$ and $\phi_s(x)=(1/\sqrt{\pi})e^{-x^2}$. The
parameters $a$ and $\gamma$ are taken to be $a =10^{-3}$ and
$\gamma=0^{-5}$. The right panel is the time-dependence of the total
flux $F_t(\eta, r)$ at $r=10^2$.}
\end{figure}

The right panel of Figure 1 is the total flux $F_t(\eta) \equiv \int
f(\eta, r, x)dx$. It shown again that total flux approaches to a
stable state with $F_t=1$ at time $\eta\geq 2000$. If the Ly$\alpha$
photon transfer is due only to spatial diffusion, the time scale of
a spatial transfer with an optical depth $\tau_0=10^2$ should be as
large as $\eta \sim \tau_0^2=10^4$. However, Figure 1 shows that photons
have already escaped from the $\tau_0=10^2$ hole within the time
$\eta\sim 2000$, which is much less than the time scale of a purely
Brownian diffusion. Therefore, the transfer of photons should not be a
process of purely Brownian diffusion. This point is very well known
in early studies on the escape of resonant photon from opaque clouds
(Osterbrock 1962; Harrington 1973; Avery \& House 1968). The time
scale of escape shown in Figure 1 is also consistent with the
estimation given by Monte Carlo simulations (Adams, 1975; Bonilha et
al. 1979). However, these works are mainly based on the time scale
of the escape of photons with frequency, at which the photon can
take a ``single longest excursion" (Adams 1972). Figure 1 shows that
the escape time scale $\eta = 2000$ is available not only for
photons which can take a ``single longest excursion", but also for
all photons with frequency around $\nu_0$ or $x=0$.

The right panel of Figure 1 also plays the role of testing our
algorithm. Because $\int\phi(x,a)jdx=\int R(x,x';a)jdxdx'$, eq.(9)
yields
\begin{equation}
F_t(\eta, r)=F_t(\eta, 0)=S_0
\end{equation}
when the solution approached the stable state. Equation (14) is the
conservation of photon number. Since numerical errors which have accumulated
over a long time evolution could be huge, eq.(14) is useful to check
the reliability of the code. The right panel of Figure 1 shows
perfectly $F_t(\eta, r=10^2) =F_t(\eta, 0)=S_0=1$ at stable state.

\subsection{Time scale of local thermalization}

Figure 2 shows a solution of the mean intensity $j(\eta,r,x)$ of
equations (9) and (10) at $r=10^2$ and time $\eta=200$, 300, 500.
The source is the same as Figure 1, namely $S_0=1$ and
$\phi_s(x)=(1/\sqrt{\pi})e^{-x^2}$. Other parameters are also the
same as Figure 1.

\begin{figure}[htb]
\begin{center}
\includegraphics[scale=0.30]{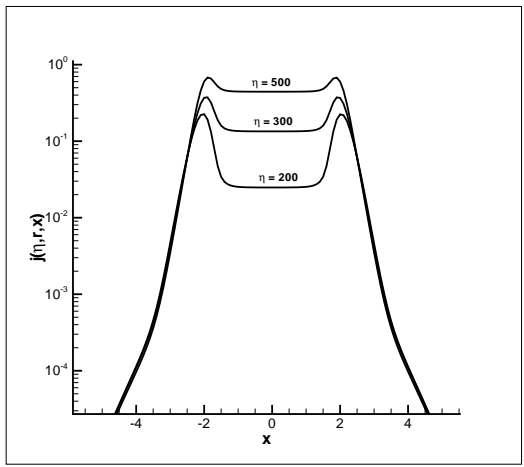}
\end{center}
\caption{Mean intensity $j(\eta,r,x)$ at $r=10^2$ of equations (9)
and (10) at time $\eta=200$, 300, 500. The source is $S_0=1$ and
$\phi_s(x)=(1/\sqrt{\pi})e^{-x^2}$. The parameters $a$ and $\gamma$
are taken to be $a =10^{-3}$ and $\gamma=0^{-5}$.}
\end{figure}

A remarkable feature of the solutions is to show a flat plateau
around $x=0$. As has been shown by Roy et al. (2009c), the flat
plateau actually is the Boltzmann statistical equilibrium
distribution around $x=0$ when the atomic mass is infinite. If the
mass is finite, i.e. considering the recoil in the re-distribution
functions (6) or (7), the flat plateau will become $e^{-2bx}$, where
$b=h\nu_0/mv_Tc$. This is the local Boltzmann distribution required
by the Wouthuysen-Field effect (Roy et al. 2009b).

Without resonant scattering, the pure absorption will lead to the
flux of $\nu_0$ photons at $r=10^2$ to be attenuated by a factor
$e^{\tau(0)}\simeq 10^{-25}$. Therefore, Figure 2 shows that a major
effect of resonant scattering is to restore $\nu_0$ photons in
optically thick halos. According to the re-distribution function
$\mathcal{R}(x,x')$ equation (5), the probability of transferring a
$x'$ photon to a $|x|<|x'|$ photon is larger than that from $x'$ to
$|x|>|x'|$. Therefore, the net effect of resonant scattering is to
bring photons with frequency $x'\neq 0$ to the central range $x \sim
0$ of frequency space. Photons of $x\sim 0$ are then effectively
restored. Moreover, the restored photons are thermalized. We see
from Figure 2 that in the time range from $\eta=200$ to $500$, the
mean intensity $j$ at $x=0$ quickly increases by a factor larger
than $10$. In the same time, the flat plateau or the local
thermalization around $x=0$ is perfectly held. Therefore, the time
scales $t_{ther}$ of $\nu_0$ photon restoring and thermalization
should be less than $\eta=200$, which is much smaller than the time
scale $t_{escape}$ of escaping from a halo of $r=10^2$.

\begin{figure}[htb]
\begin{center}
\includegraphics[scale=0.35]{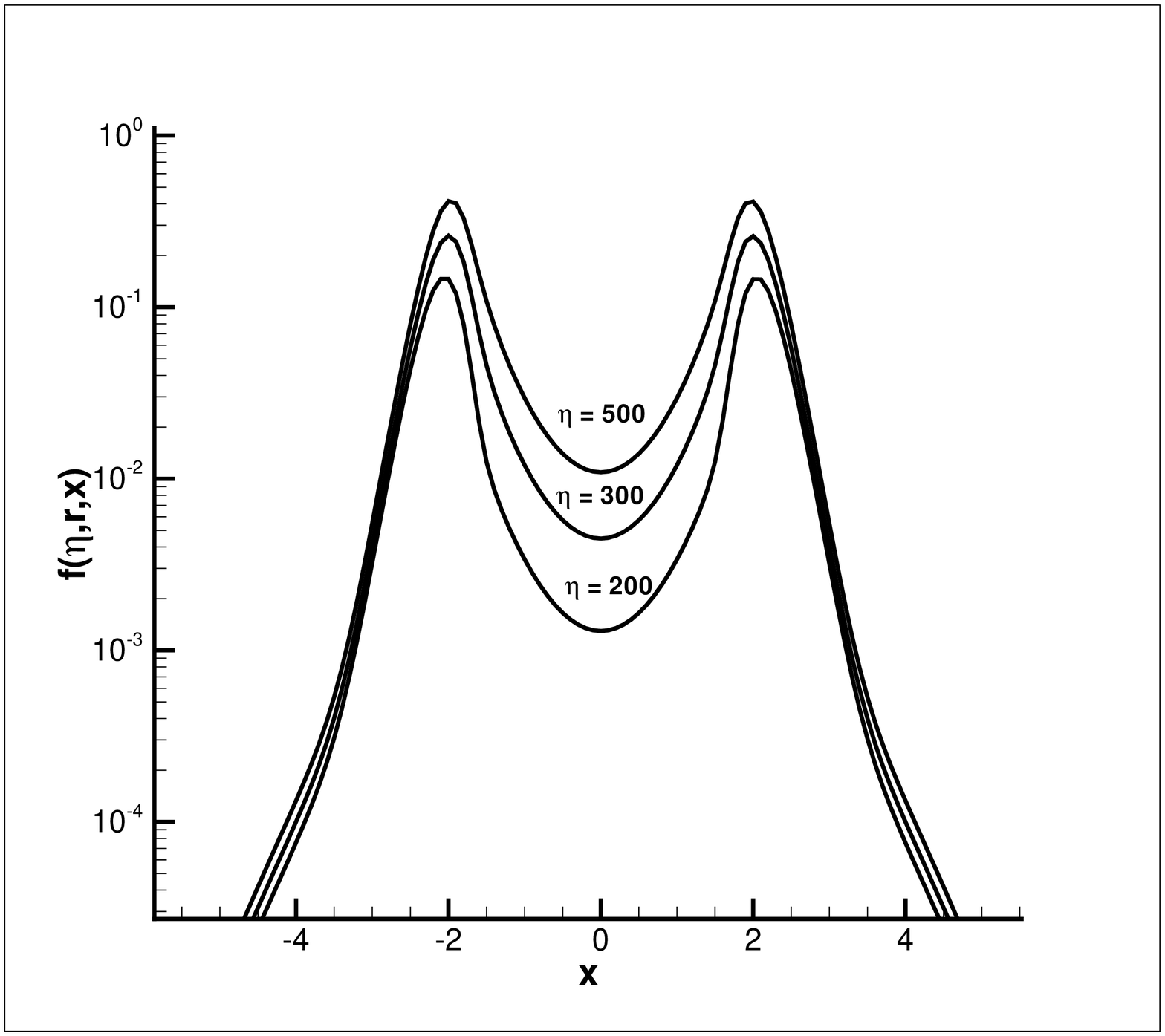}
\includegraphics[scale=0.35]{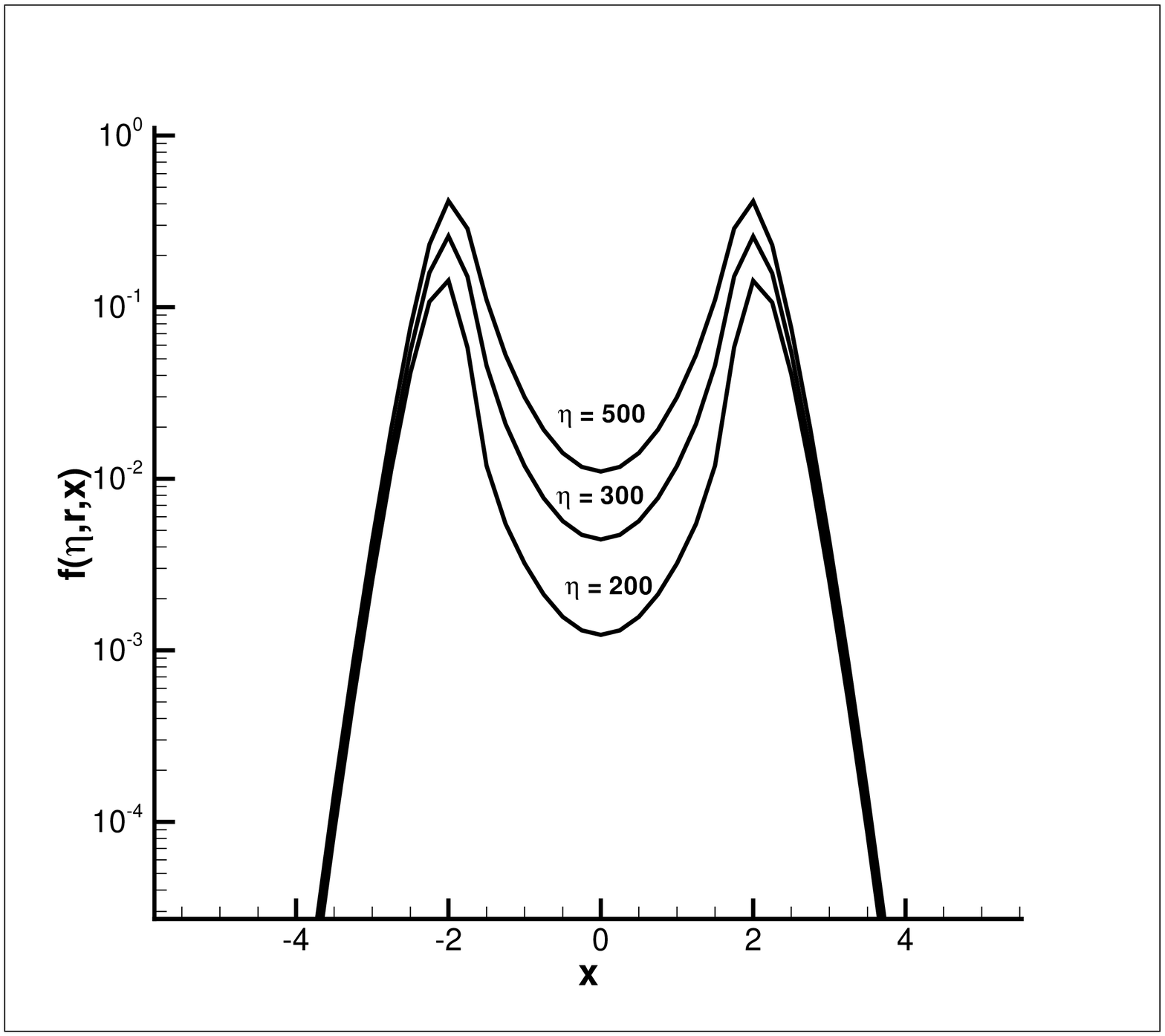}
\caption{Flux $f(\eta,r,x)$ at $r=10^2$ and time $\eta=200$, 300 and
500, with sources left panel: $S_0=1$ and
$\phi_s(x)=(1/\sqrt{\pi})e^{-x^2}$, and right panel: $S_0=1$ and
$\phi_s(x)=(1/\sqrt{\pi/2})e^{-2x^2}$. Other parameters are the same
as in Figure 1}
\end{center}
\end{figure}

The result of $t_{ther}\ll t_{escape}$ in optically thick halos is
very important. One can conclude that the photons of the flux $f$
around $x=0$ emergent from an optically thick halos actually come
from the thermalized photons, regardless of the initial distribution of
the photons. The initial conditions of the photon source have already
been forgotten during the thermalization. Therefore, one can expect
that the profile of flux $f$ has to be independent of the sources. We
demonstrate this point with Figure 3.

Figure 3 presents the flux $f$ given by equations (9) and (10) with
the same parameters as the solution of Figure 1, and the source
profiles are  $\phi_s(x)=(1/\sqrt{\pi})e^{-x^2}$ (left) and
$\phi_s(x)=(1/\sqrt{\pi/2})e^{-2x^2}$ (right). That is, the line
widths are, respectively,  1  and $1/\sqrt{2}$. We see that the left
and right profiles of Figure 3 are almost identical within $|x|\leq
3$. Therefore, the initial distribution of photon frequency is
already forgotten, and the photons of the left and right cases at
$\eta>200$ actually are from the same thermalized sources. The
profiles of the flux are also held if the line width is broader than
1. We will show this point when the source has a continuant spectrum
(\S 3.4).

\subsection{Two peaks in the flux profile}

The profiles of either $j$ or $f$ shown in Figures 1 - 3 have two
peak structure. The flux $f$ is dominated by photons with frequency
$x_{\pm}\simeq \pm (2-3)$. The two peak structure has been
recognized in all the time-independent solutions of the
Fokker-Planck approximation (Harrington 1973; Neufeld 1990; Dijkstra
et al. 2006), and Monte Carlo simulations (Lee 1974; Zheng \&
Miralda-Escude 2002; Ahn et al. 2002; Cantalupo et al. 2005;
Verhamme et al. 2006).  A point we would like to emphasize is that
this structure is independent of the profile of the source $S$. It
is because the initial properties of the central sources have been
forgotten during the local thermalization.

Since the shape of the locally thermalized distribution is time-independent, the
frequency of the two peaks, $|x_{\pm}|$, at a given $r$ is also
time-independent. When the photons of the flux $f$ mainly come from the
locally thermalized photons, the frequency of the two
peaks, $|x_{\pm}|$, should not be described by the relation
$|x_{\pm}|= (a\tau)^{1/3}$, because this relation is from a solution
of the time-independent Fokker-Planck equation, which does not describe
the thermalization (Neufeld 1990).

\begin{figure}[htb]
\begin{center}
\includegraphics[scale=0.35]{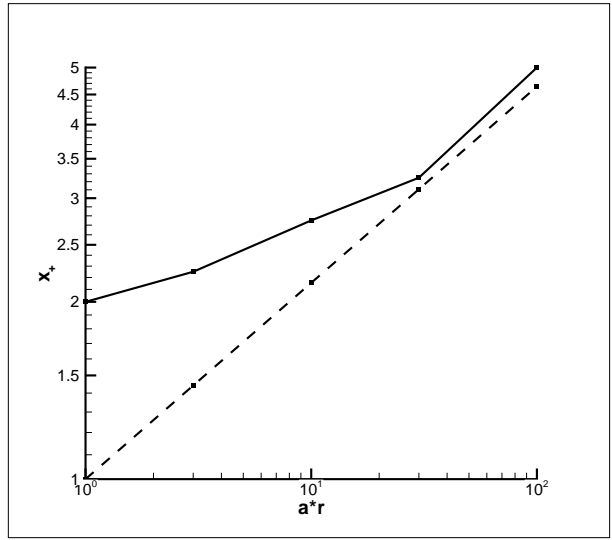}
\caption{The positions of the peaks $|x_{+}|$ as a function of
$(ar)^{1/3}$ for the solutions of eqs. (9) and (10) with $ a = 10^{-2}$.
Other parameters are the same as in Figure 1. The dashed line is for
$x_{\pm} =\pm (ar)^{1/3}$.}
\end{center}
\end{figure}

Figure 4 presents the peak frequency $|x_{\pm}|$ as a function of
$ar$ for solutions given by equations (9) and (10) with $a =
10^{-2}$. We consider only $r\geq 10^2$, as in the case $r\leq 10^2$
photons do not undergo a large enough number of scattering, and
therefore, are not completely thermalized yet. With the
dimensionless variables, $ar$ is equal to $a\tau$. A line
$|x_{\pm}|= (ar)^{1/3}$ is also shown in Figure 4. It shows that our
numerical result of $|x_{\pm}|$ is significantly different from the
$(a\tau_0)^{1/3}$-law if $ar < 30$. That is, in the range $ar < 30$,
photons of the flux $f$ are dominated by the locally thermalized
photons.  The frequency $x_{\pm}$ actually is dependent on the width
of the flat plateau or the locally thermalized distribution of $j$.
Therefore, $x_{\pm}$ is always larger than two. This feature has
also been addressed in Bonilha et al (1979). The curve of Figure 4
is approximately available for $a=10^{-3}$ and $10^{-4}$. Thus,
$|x_{\pm}|$ is larger than $(ar)^{1/3}$ if $r < 3\times 10^5$ and
$3\times 10^6$ for $a=10^{-3}$ and $10^{-4}$, respectively.

Figure 4 shows a slowly increase of $|x_{\pm}|$ with $r$ in the
range $ar\leq 30$, and then, it approaches $(ar)^{1/3}$ for
larger $ar$. When $r$ is large, more photons of the flux are attributed
to the resonant scattering by the Lorentzian wing of the Voigt function.
$|x_{\pm}|$ is then approaching to $(ar)^{1/3}$. It should
be emphasized once again that all these results are independent of
the intrinsic width of Ly$\alpha$ emission from the central source.

\subsection{Absorption spectrum}

If the radiation from the sources has a continuant spectrum, the
effect of neutral hydrogen halos is to produce an absorption line at
$\nu=\nu_0$. The profile of the absorption line can also be found by
solving equations (9) and (10), but replacing the boundary equation
(11) by
\begin{equation}
j(\eta, 0, x)=0, \hspace{1cm} f(\eta, 0, x)=S_0.
\end{equation}
That is, we assume that the original spectrum is flat in the frequency space.

\begin{figure}[htb]
\begin{center}
\includegraphics[scale=0.30]{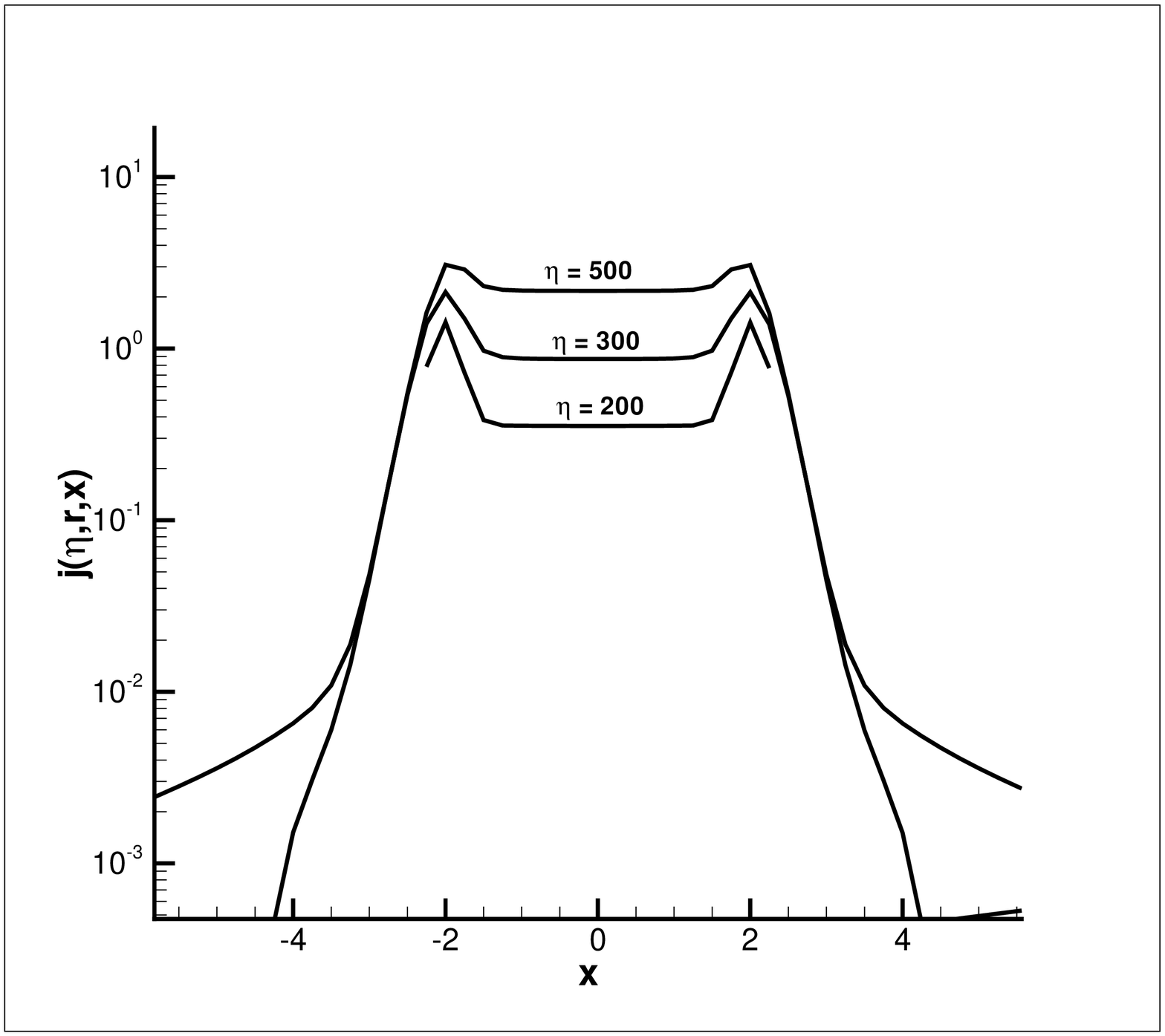}
\includegraphics[scale=0.30]{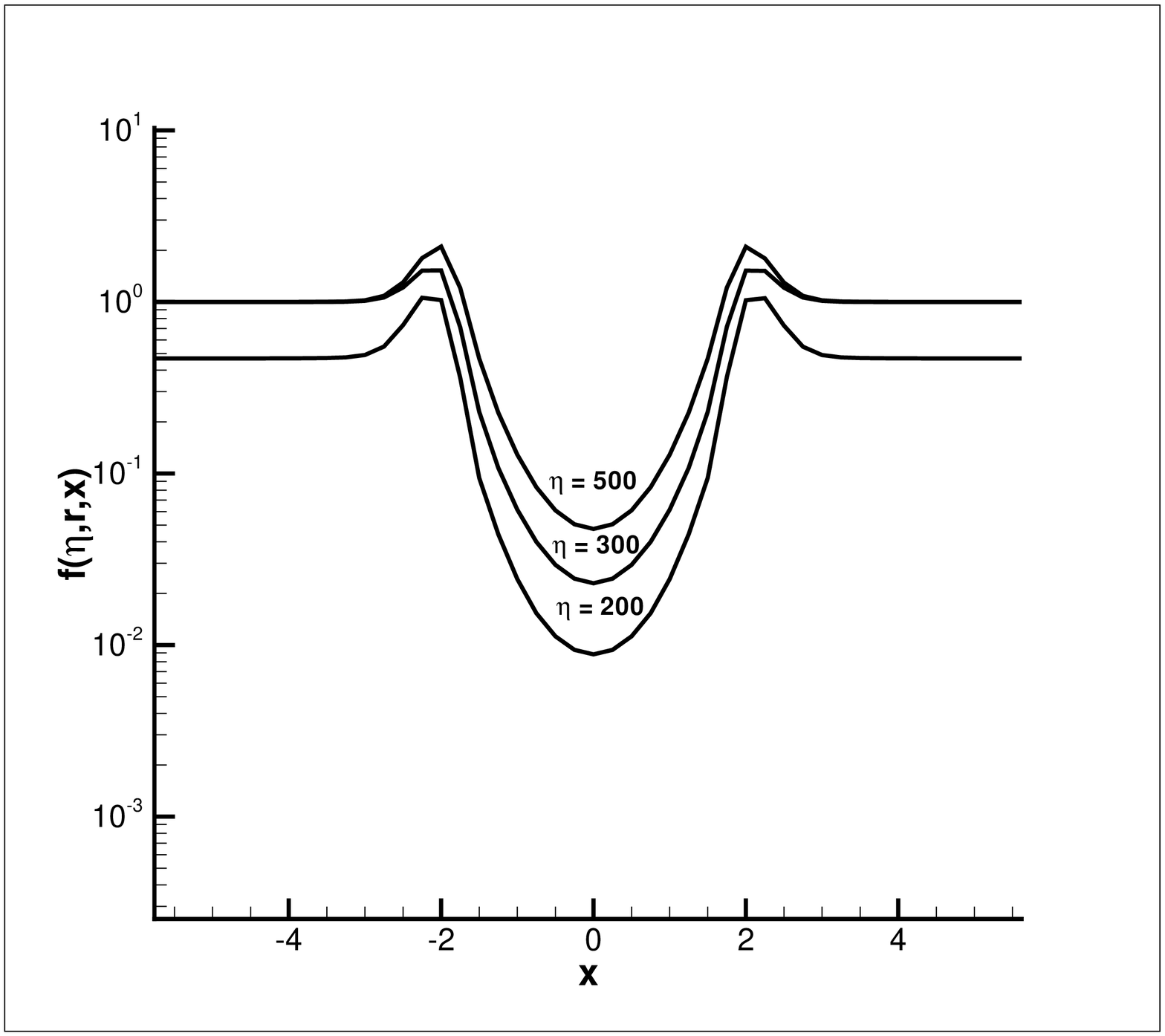}
\caption{Solutions of $j(\eta, r, x)$ (left)  and $f(\eta, r, x)$
(right) of eqs. (9) and (10) at $r=10^2$ and $\eta=200$, $ 300$ and
$500$. The source is given by eq. (16). The parameters are $a
=10^{-3}$ and $\gamma=10^{-5}$. }
\end{center}
\end{figure}

A solution of the time evolution of $j$ and $f$ at $r=10^2$ with the
source equation (15) is shown in Figure 5. The optical depths at the
frequency $|x| >4$ are small, and therefore, the Eddington
approximation would no longer be proper. However, those photons do
not strongly involve the resonant scattering, and hence they do not
significantly affect the solution around $x=0$. Therefore, the
solution is still useful to study the profiles of $j$ and $f$ around
$x=0$. The profile of $f$ typically is an absorption line with width
given by the position of the two peak structure. As expected, the
profile in the range $|x|<4$ is the same as the left panel
of Figure 1. It shows again that the two peak structure is
independent of the frequency spectrum of the central source.

The flux $f$ of Figure 5 has a flat wing in both sides of $x>3$ and
$x<-3$, because the effect of resonant scattering is negligible for
photons with frequency $|x|>4$, and they can freely transfer from
$r=0$ to $10^2$. On the other hand, the mean intensity $j$, which is
the isotropic component of the intensity $J$ [eq.(8)], does not have
wings at $|x|>4$. That is, resonant scattering cannot store photons
with $|x|>4$ in the halo. Nevertheless, the mean intensity $j$ still has
a flat plateau around $x=0$. It means that in the period from
$\eta=200$ to $\eta=500$, the halo trapped and stored more and more
photons of $|x|<4$. These photons are in the locally thermalized
state.

\subsection{The estimation of the HI column density}

As an application of the absorption spectrum in Figure 5, we study the
profile of the red damping wing of the optically thick IGM at high
redshifts. Since the variable $x$ is independent of redshift, the
profile of a red damping wing is directly given by the flux $f(\eta,
r, x)$ at the wing of low frequency $x\leq 0$.

\begin{figure}[htb]
\begin{center}
\includegraphics[scale=0.35]{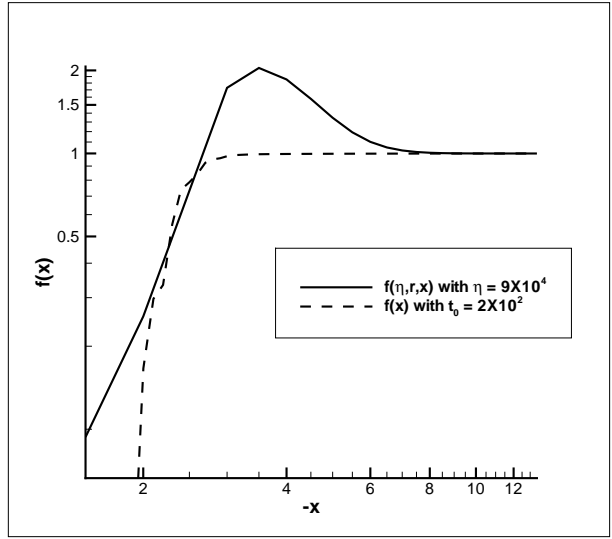}
\caption{Red damping wing of (a) the DLA model eq.(16)
$\tau_0=2\times 10^2$ (dashed line);  (b) the solution $f(x)$ of \S
3.4 at $r=10^4$ and $\eta=9\times 10^4$ (solid line). For both (a)
and (b), the parameter $a =10^{-2}$.}
\end{center}
\end{figure}

If the hydrogen clouds are located far from the Ly$\alpha$ sources,
the red damping wing can be modeled as the absorption of an
optically thick halos. The red damping wing of damped Ly$\alpha$
system (DLA) model is then given by $f(x)=e^{-\tau(x)}$ and $x<0$,
where $\tau(x)$ is from the Voigt function equation (2) as
\begin{equation}
\tau(x)=\tau_0\phi(x,a)=\tau_0\frac
{a}{\pi^{3/2}}\int^{\infty}_{-\infty}
dy\frac{e^{-y^2}}{(y-x)^2+a^2}.
\end{equation}
The column number density of neutral hydrogen atoms, $N_{\rm HI}$,
generally is estimated with fitting profile equation (16) with
observation, and then, $N_{\rm HI}$ can be found from $\tau_0$ by
equation (4). If the light source is located in a hydrogen cloud,
the column number density given by the fitting of equation (16)
should be underestimated, because resonant scattering helps the
transfer of resonant photons.

As an example Figure 6 gives $(a)$ the red damping wings of eq.(16)
with $\tau_0=2\times 10^2$, and $(b)$ a solution $f$ of \S 3.4 at
$r=10^4$, or $\tau_0=10^4$ and $\eta=9\times 10^4$. The profiles of
$(a)$ and $(b)$ are very different from each other. The former is
quickly dropping when $|x|$ is less than about 3, while the later at
$|x|<3$ is softly dependent on $x$. The curve of $(b)$ approaches to
a much higher bottom at $x=0$.

More importantly, Figure 6 shows that the curve $(a)$ with
$\tau_0=2\times 10^2$ is more or less close to the curve $(b)$. That
is, the DLA model at $\tau_0=2\times 10^2$ may give a similar data
fitting as the solution with resonant scattering. Therefore, for
optical depth $\tau_0=10^4$, the DLA model of equation (16) will
underestimate the column number density of neutral hydrogen by about
two orders.

\section{Solutions of flash sources}

\subsection{Frequency-dependence of Ly-alpha transfer}

If the light source is significantly time-dependent, like the
optical afterglow of GRBs, we can treat the source as a flash described by a
boundary conditions as:
\begin{equation}
j(\eta, 0, x)=0, \hspace{1cm} f(\eta, 0, x) =
    \left \{ \begin{array}{ll}
        S_0 \phi_s(x), & \mbox{$0<\eta <\eta_0$} \\
        0, & \mbox{$\eta>\eta_0$}.
       \end{array} \right.
\end{equation}
It describes a flash within a time range $0<\eta < \eta_0$, or
the time duration is $\Delta\eta=\eta_0$. We consider the case of $r
\geq \eta_0$. That is, the size of the halo is larger than the
spatial range lasted by the flash, as photon spatial transfer in
optically thick medium cannot be faster than the speed of light. The
initial condition is still the same equation (13).

\begin{figure}[htb]
\begin{center}
\includegraphics[scale=0.28]{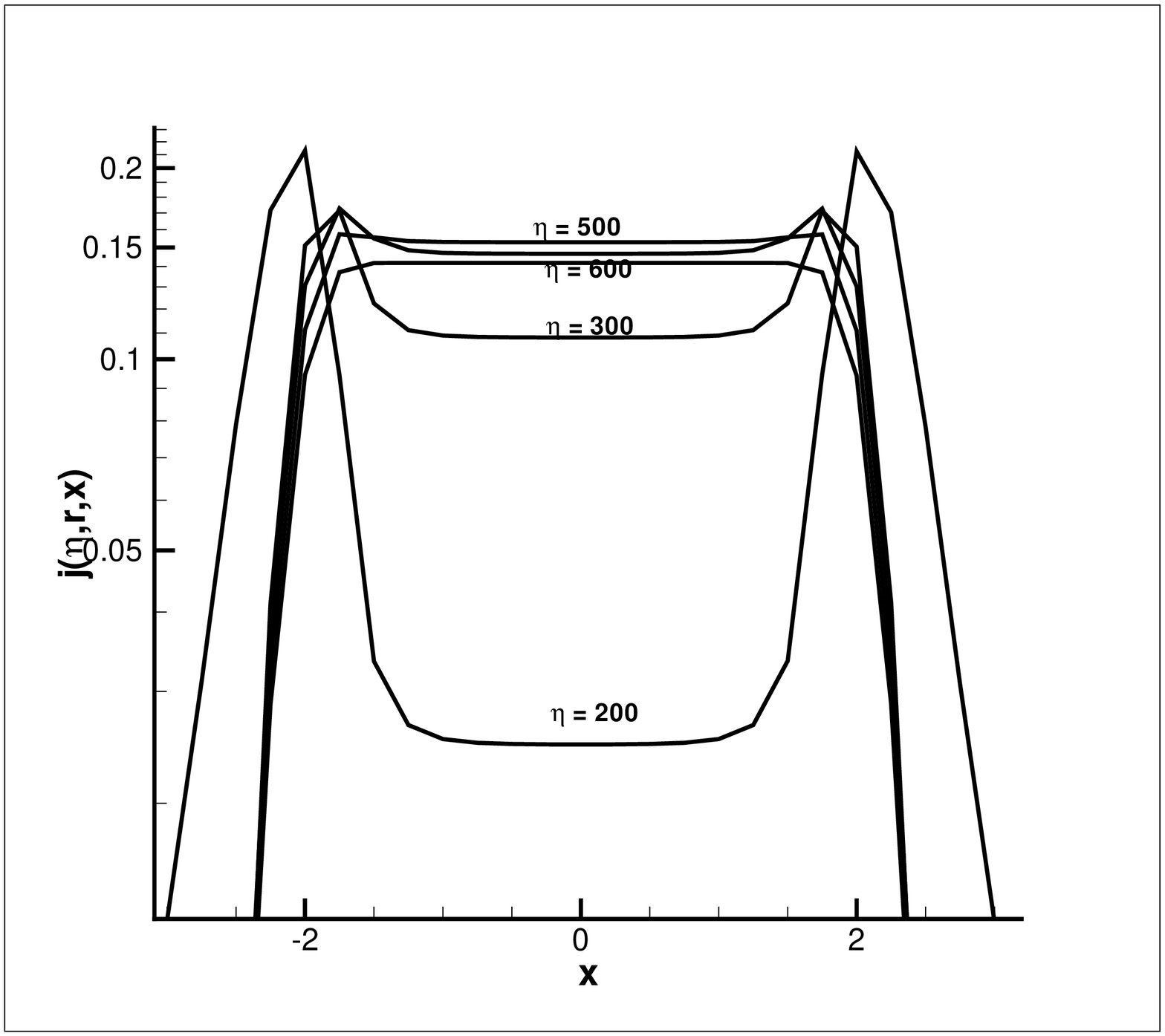}
\includegraphics[scale=0.28]{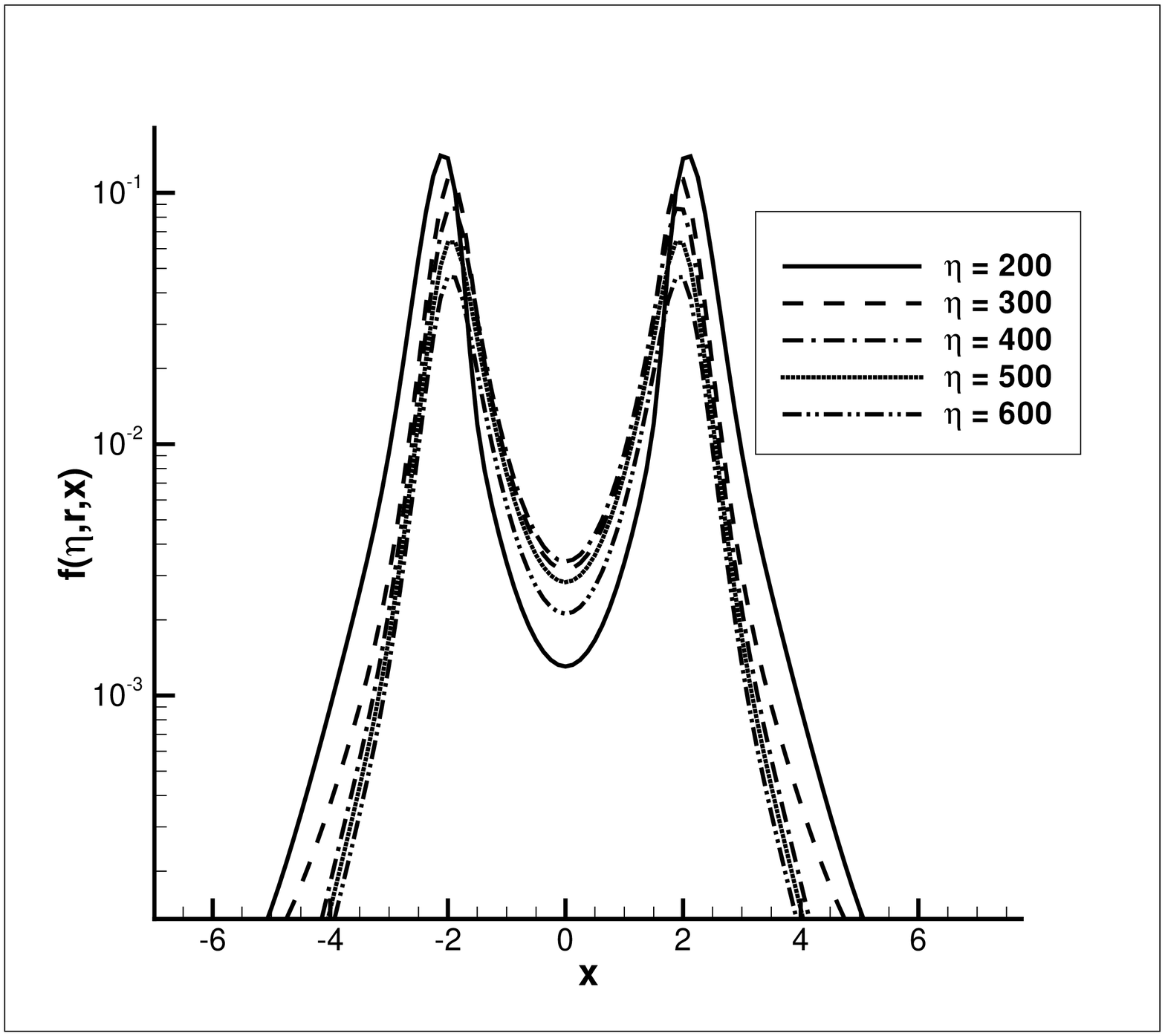}
\includegraphics[scale=0.28]{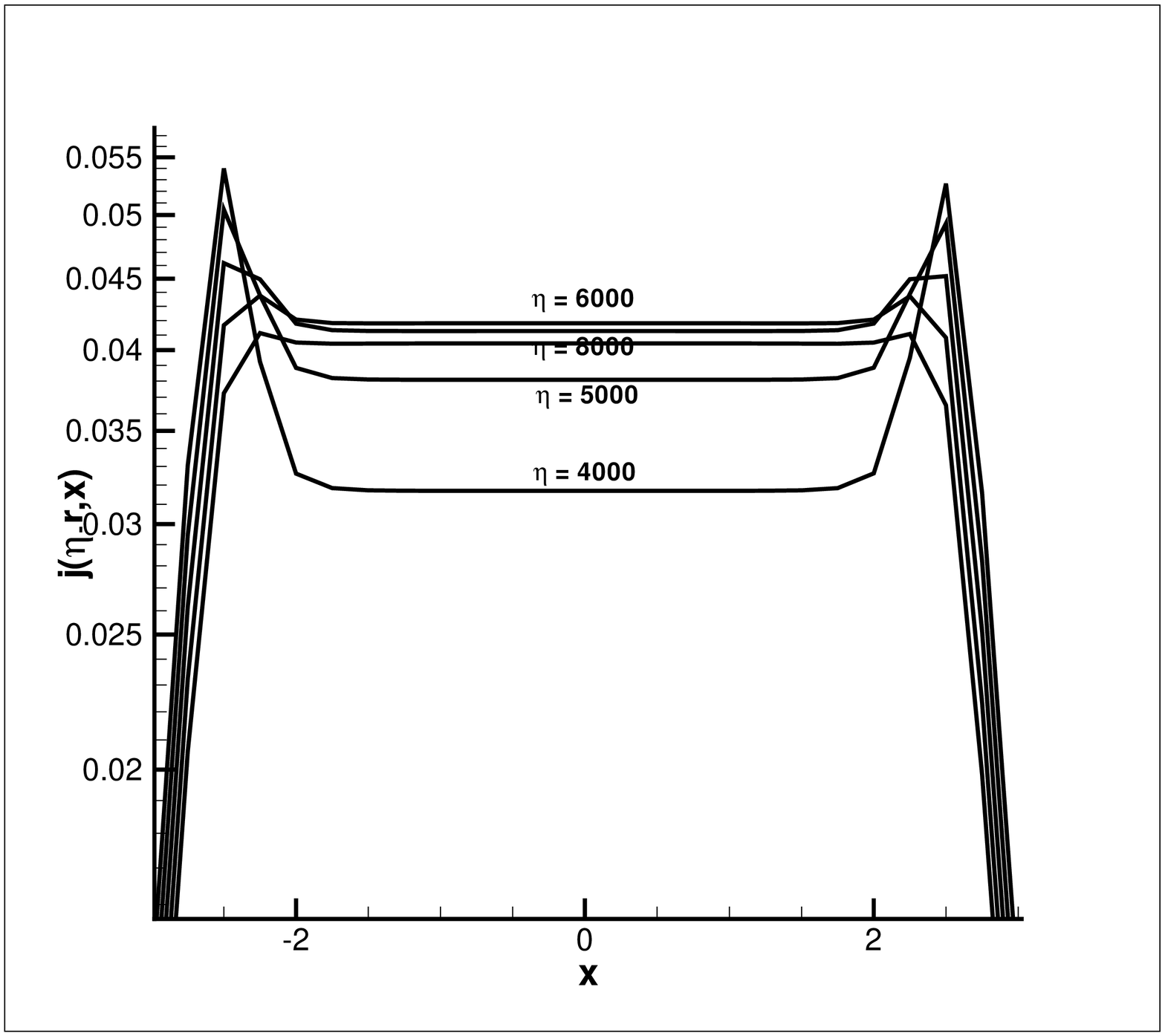}
\includegraphics[scale=0.28]{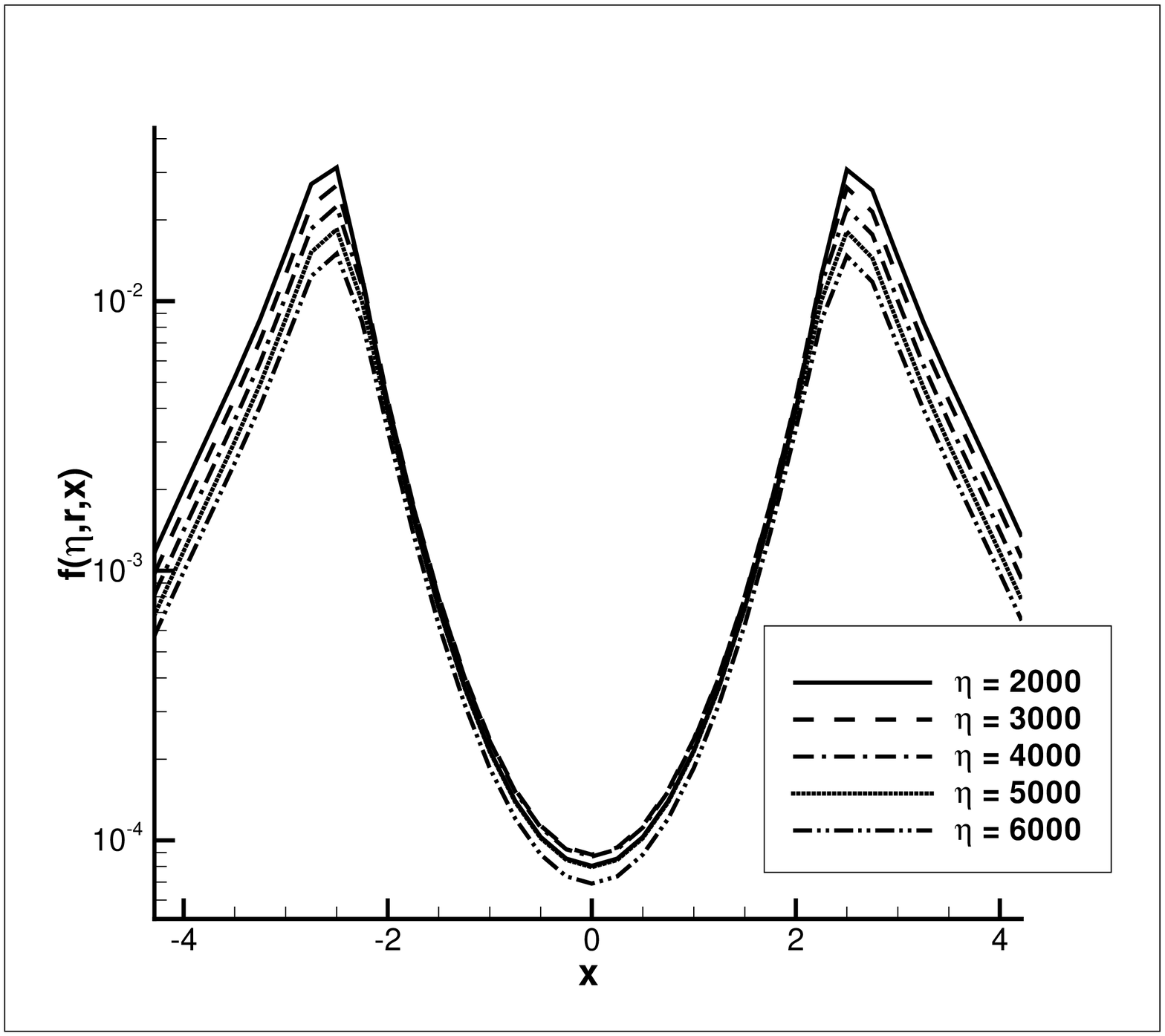}
\caption{The profiles of $j(\eta, r,x)$  and $f(\eta,r,x)$ with
time-dependent source [eq.(17)]. Top panel: $j(\eta, r, x)$ (left)
and $f(\eta, r, x)$ (right) of $\eta_0=100$ at $r=10^2$. Bottom
panel:  $\eta_0 = 500$ at $r=10^3$. The parameters $a =10^{-2}$ and
$\gamma=10^{-5}$.}
\end{center}
\end{figure}

Figure 7 presents two time-dependent solutions of the mean intensity
$j$ and flux $f$: one is for a flash of equation (17) with $\eta_0=100$ at
$r=10^2$; the other is for a flash with $\eta_0=500$ at $r=10^3$.
Other parameters are $S_0=1$, $a=10^{-2}$ and
$\phi_s(x)=(1/\sqrt{\pi}) \, e^{-x^2}$. We still see the typical
flat plateau of $j$ in all times, even when the original time duration of the
flash is as short as $\Delta \eta=100$.

The time dependence of $j$ has a rising phase and a decaying phase.
The thermalization of $j$ is held in both rising and decaying
phases. We see from Figure 7 that the rising and decaying phases are
frequency-dependent. Photons at the two peaks reach their maximum
at about $\eta=200$ (top right panel) and $\eta=4000$ (bottom right panel),
while the flat plateau reaches their maximum at about $\eta=500$
(top right panel) and $\eta=6000$ (bottom right panel). That is, the halo holds a
locally thermalized photons for a much longer time than the original
time durations $\Delta \eta =100$ (top) and $\Delta \eta=500$ (bottom).

The time-evolution of $f$ also consists of a rising phase and a
decaying phase. Therefore, the flux emergent from the halo is also a flash.
However, the time duration is very different from the original one.
For the top panel, we see that the profile of $f$ is almost
time-independent from $\eta=300$ to 500. That is, the time duration
$500-300=200$ is much larger than the original one $\Delta \eta=100$.
For the bottom panel, the time-independent flux is from $\eta=3000$
to 5000, and therefore, the time duration of the flash is about
2000, which is also much larger than the original time duration
$\eta=500$.

\subsection{The light curves of a flash}

To further study the feature of time dependence of flash, we plot
Figure 8, which gives the light curve of the flux $f$ at $x=0$ (top
panels), and the total flux (bottom panels). The top panels show the
light curve of rising and decaying phases of the $\nu_0$ photon flux
$f$. With these light curves, one can find the maximum of the flux
$f$ at time $\eta_{max}$; the rising phase is then
$\eta<\eta_{max}$, and decaying phase is $\eta>\eta_{max}$. For each
curve, one can also define a time duration $\Delta \eta =
\eta_2-\eta_1$, where $\eta_1<\eta_{max}$ and $\eta_2>\eta_{max}$,
and both are given by the condition $f(\eta_{1,2}, r,0)=0.9
f(\eta_{max}, r,0)$. The time duration is then $\Delta \eta =
\eta_2-\eta_1$.

\begin{figure}[htb]
\begin{center}
\includegraphics[scale=0.26]{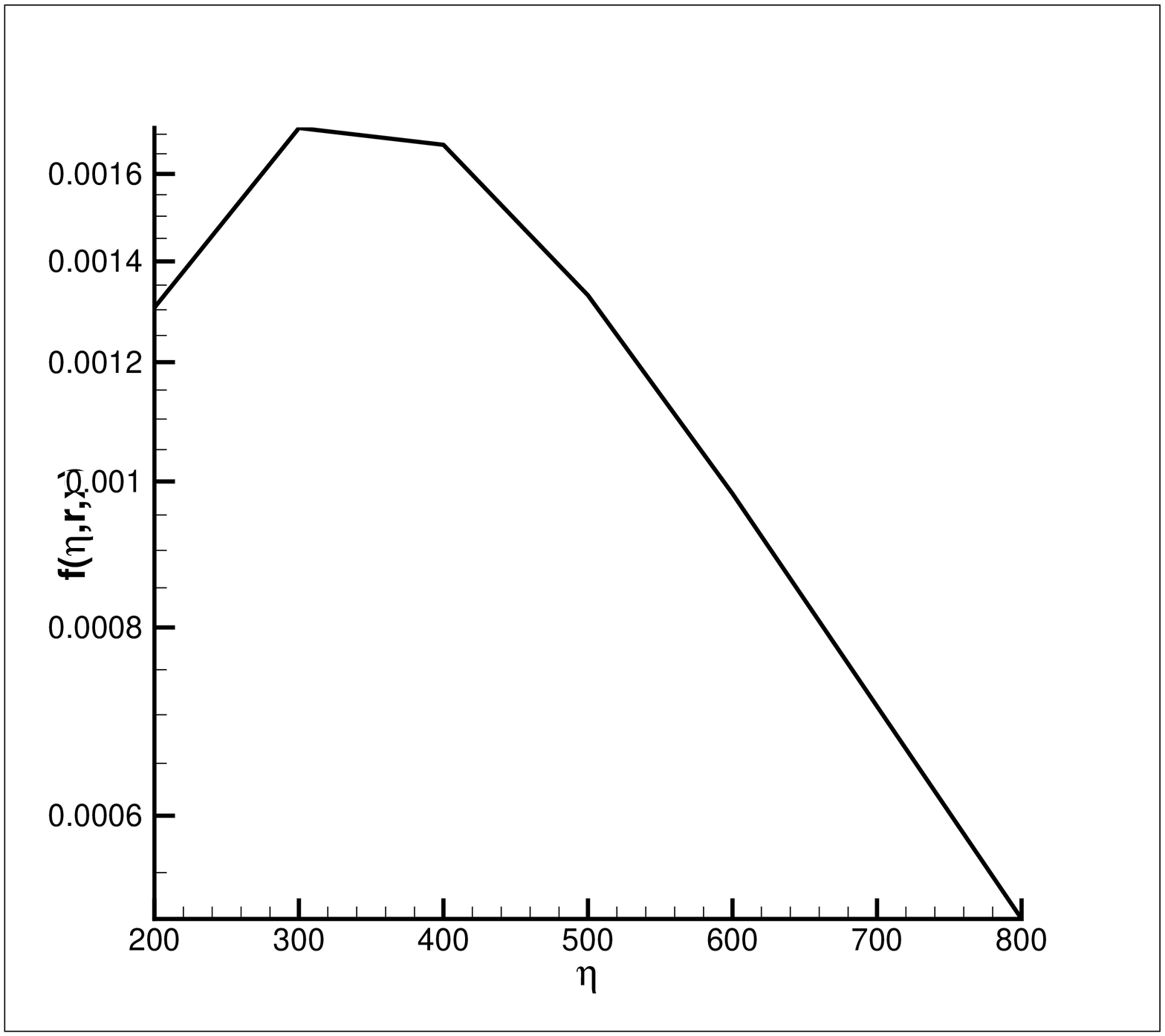}
\includegraphics[scale=0.26]{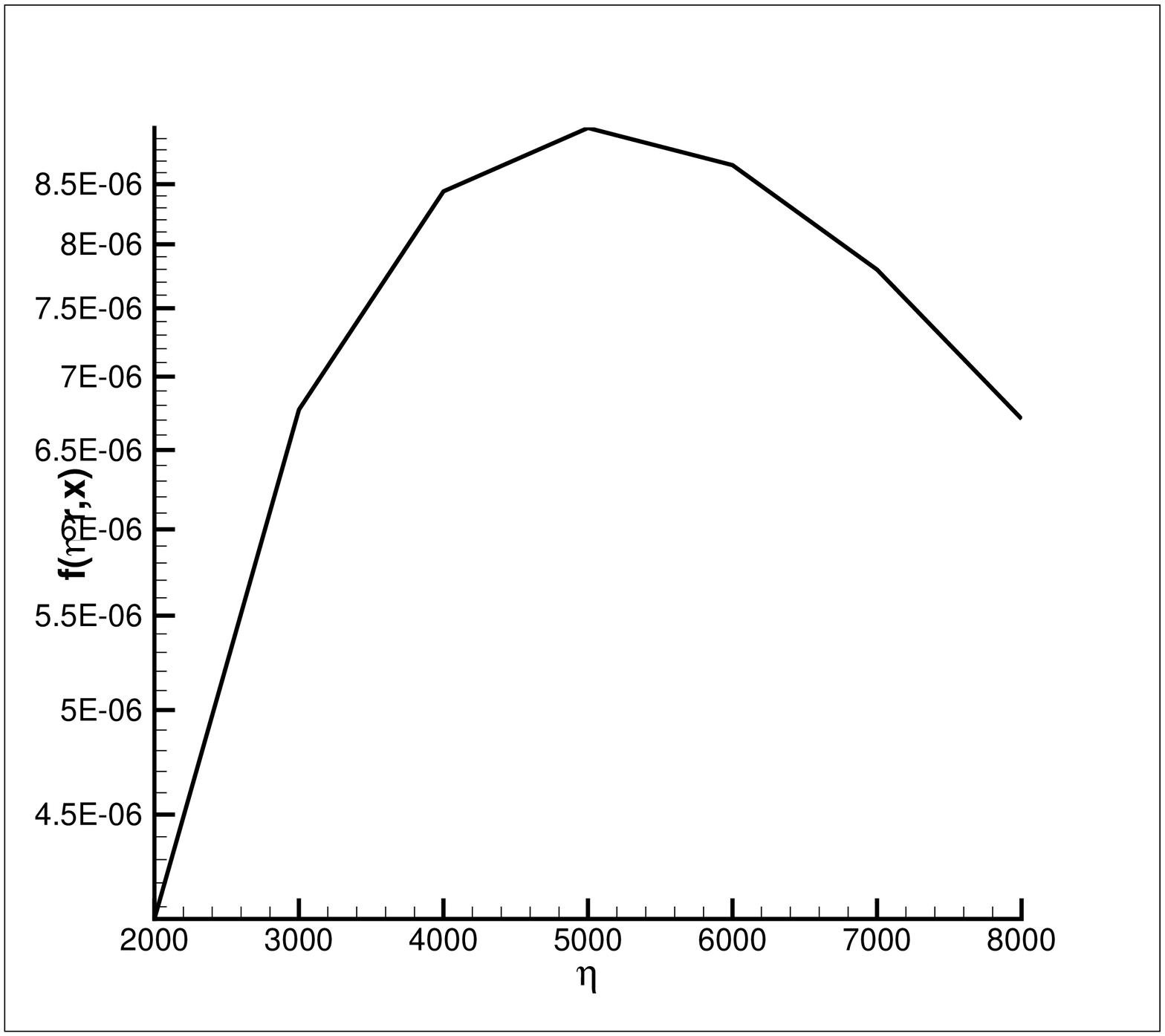}
\includegraphics[scale=0.26]{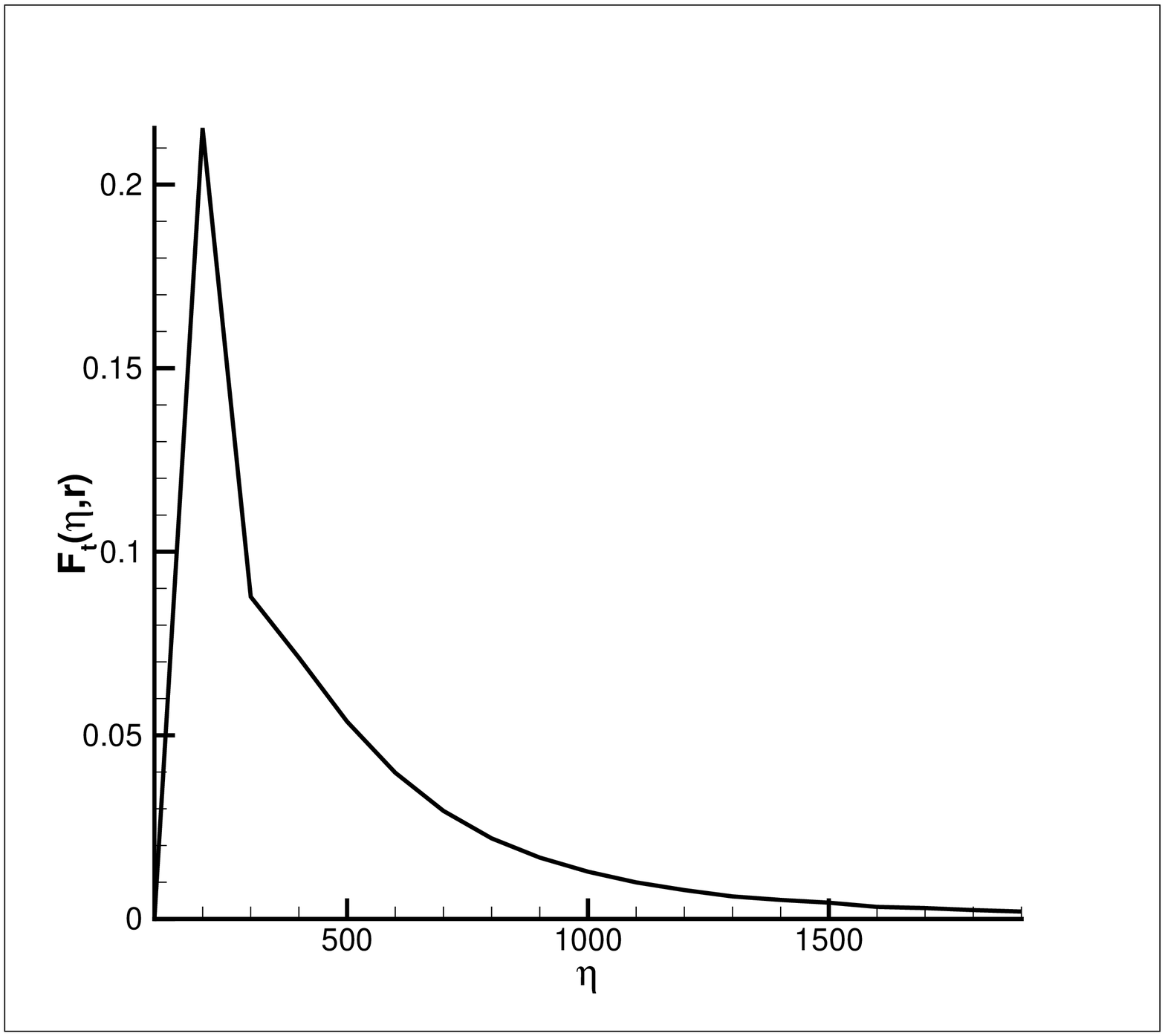}
\includegraphics[scale=0.26]{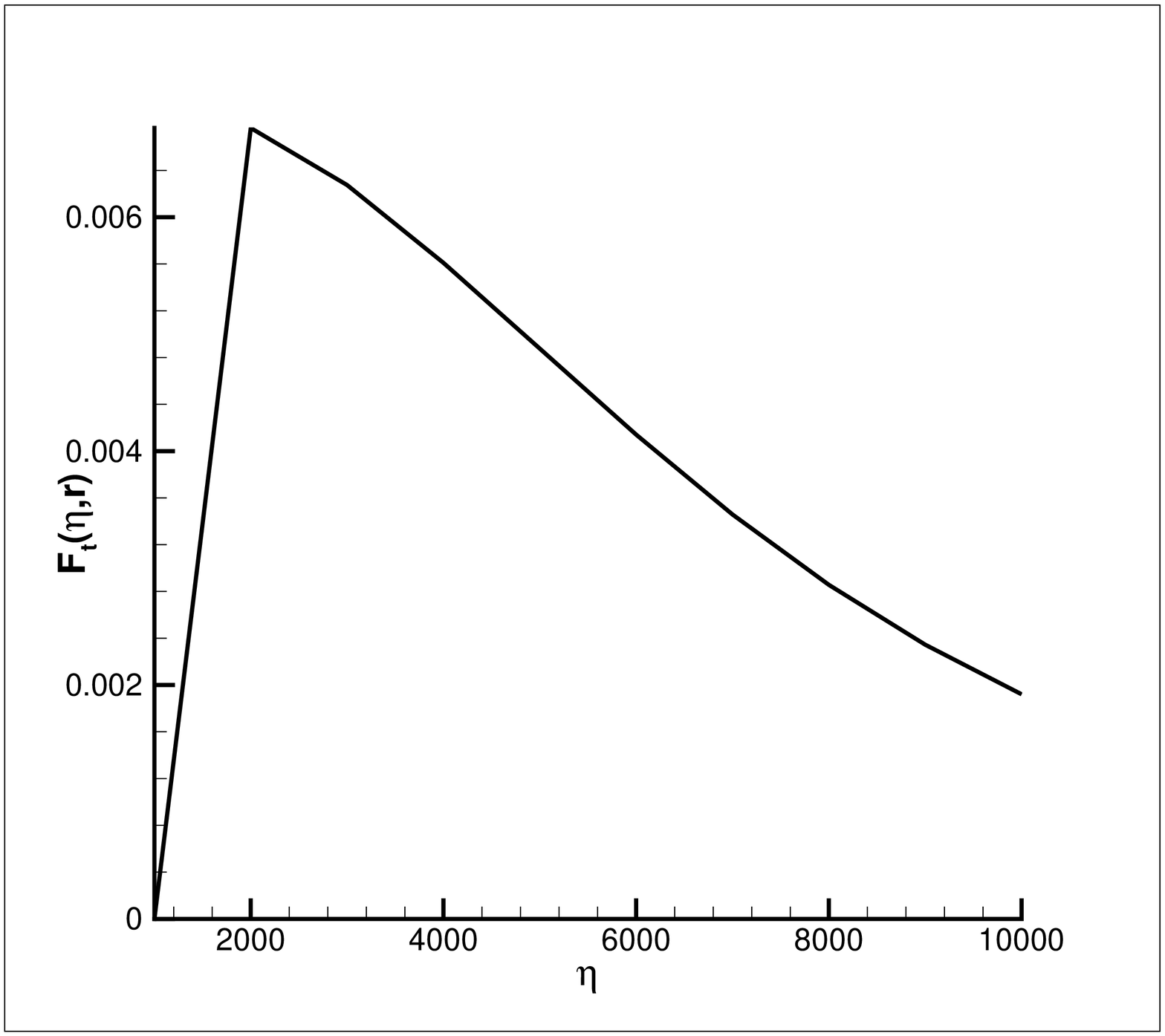}
\caption{Top panel: the light curves of $f(\eta, r, x)$ at $x=0$ at
$r=10^2$ (left) and $r=10^3$ (right) for flash source with
$\eta_0=50$. Bottom panel: the light curves of the total flux
$F_t(\eta, r)=\int f(\eta, r, x)dx$ at $r=10^2$ (left) and $r=10^3$
(right).}
\end{center}
\end{figure}

The two top panels of Figure 8 are for a flash source with original
time duration $\Delta \eta=50$. Figure 8 shows that the time
duration of the flash, $\Delta \eta$, is
$r$-dependent. At $r=0$, i.e. at the source, $\Delta \eta$ is 50.
At $r=10^2$ (top left of Figure 8), $\Delta \eta$ is about 200,
while at $r=10^3$, we have $\Delta \eta  \simeq 2000$. Therefore, the time
duration $\Delta \eta$ is increasing with $r$. This result shows
that the time duration seems to be dependent mainly on the size $r$
(or optical depth $\tau$) of the halo, regardless the initial time
duration $\Delta \eta=\eta_0$.

The bottom two panels of Figure 8 are the light curves of the total
flux $F_t(\eta, r)=\int f(\eta, r, x)dx$. The peak times
$\eta_{max}$ of total flux generally are less than that of $\nu_0$
photon flux, as the $\nu_0$ photon needs longer restoration time.
The time durations given by the total flux are also a little less
than that of $\nu_0$ photon flux, but it still shows that the time
duration is mainly dependent on the size $r$ of the halo.

\begin{figure}[htb]
\begin{center}
\includegraphics[scale=0.21]{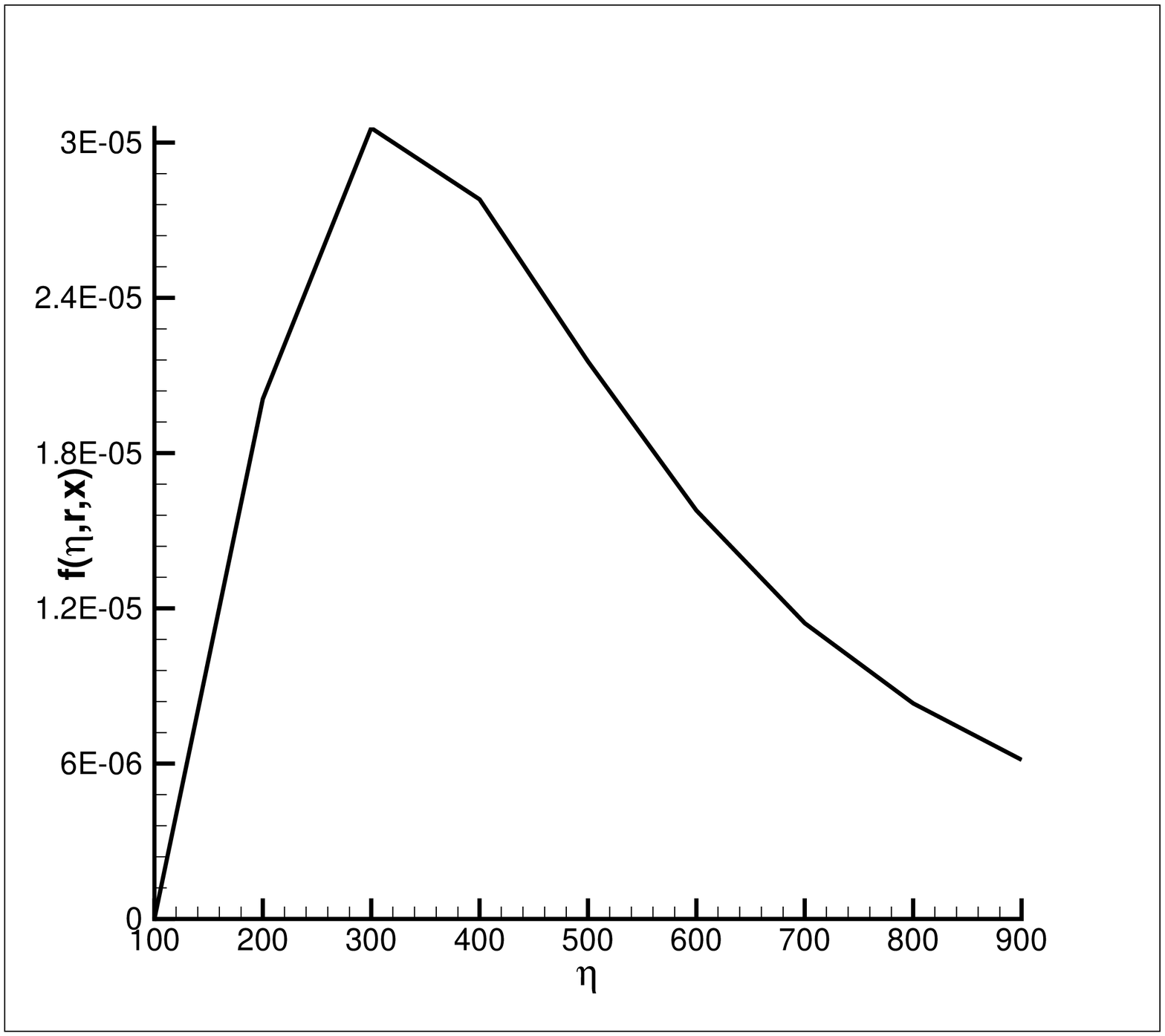}
\includegraphics[scale=0.21]{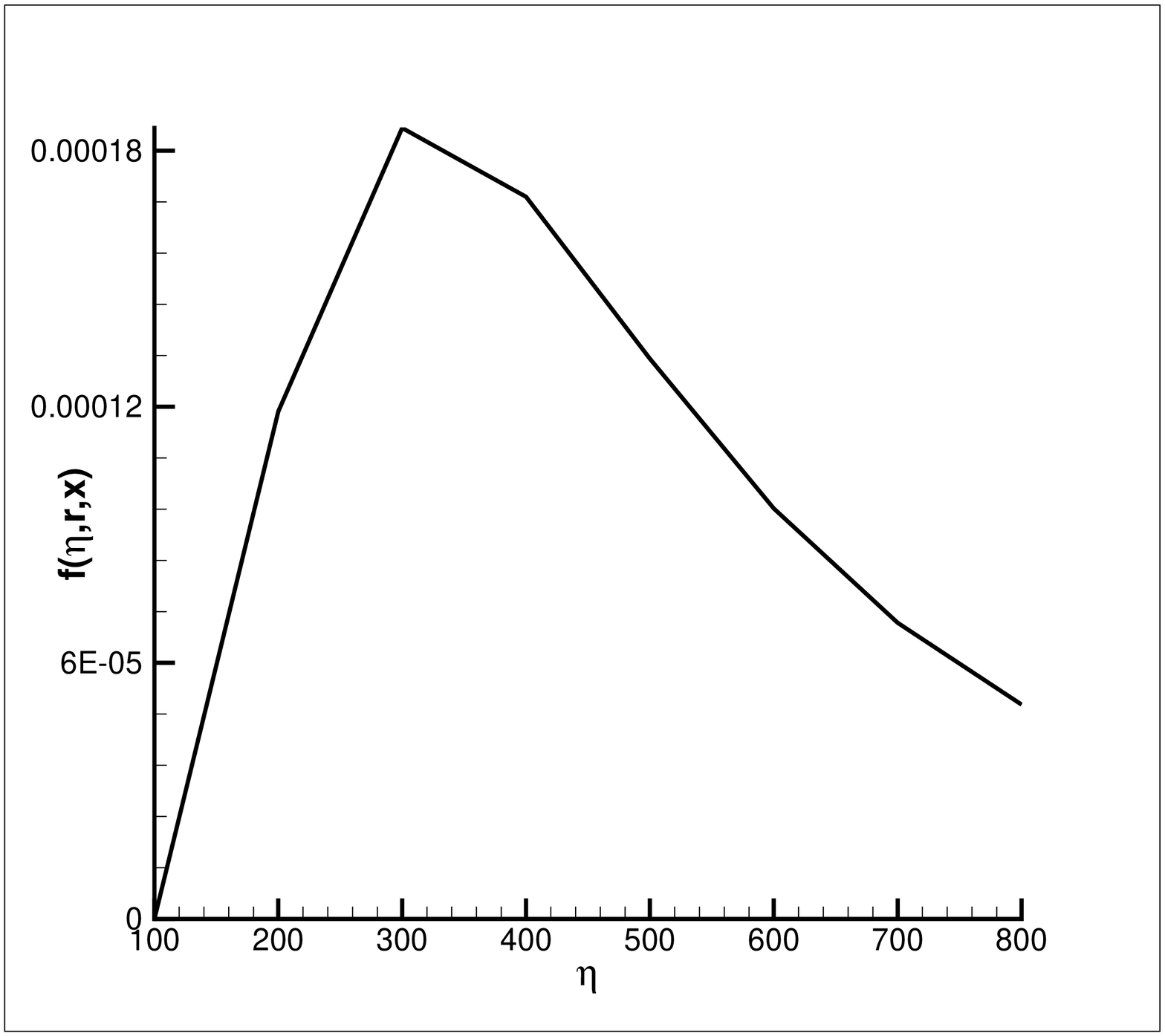}
\includegraphics[scale=0.21]{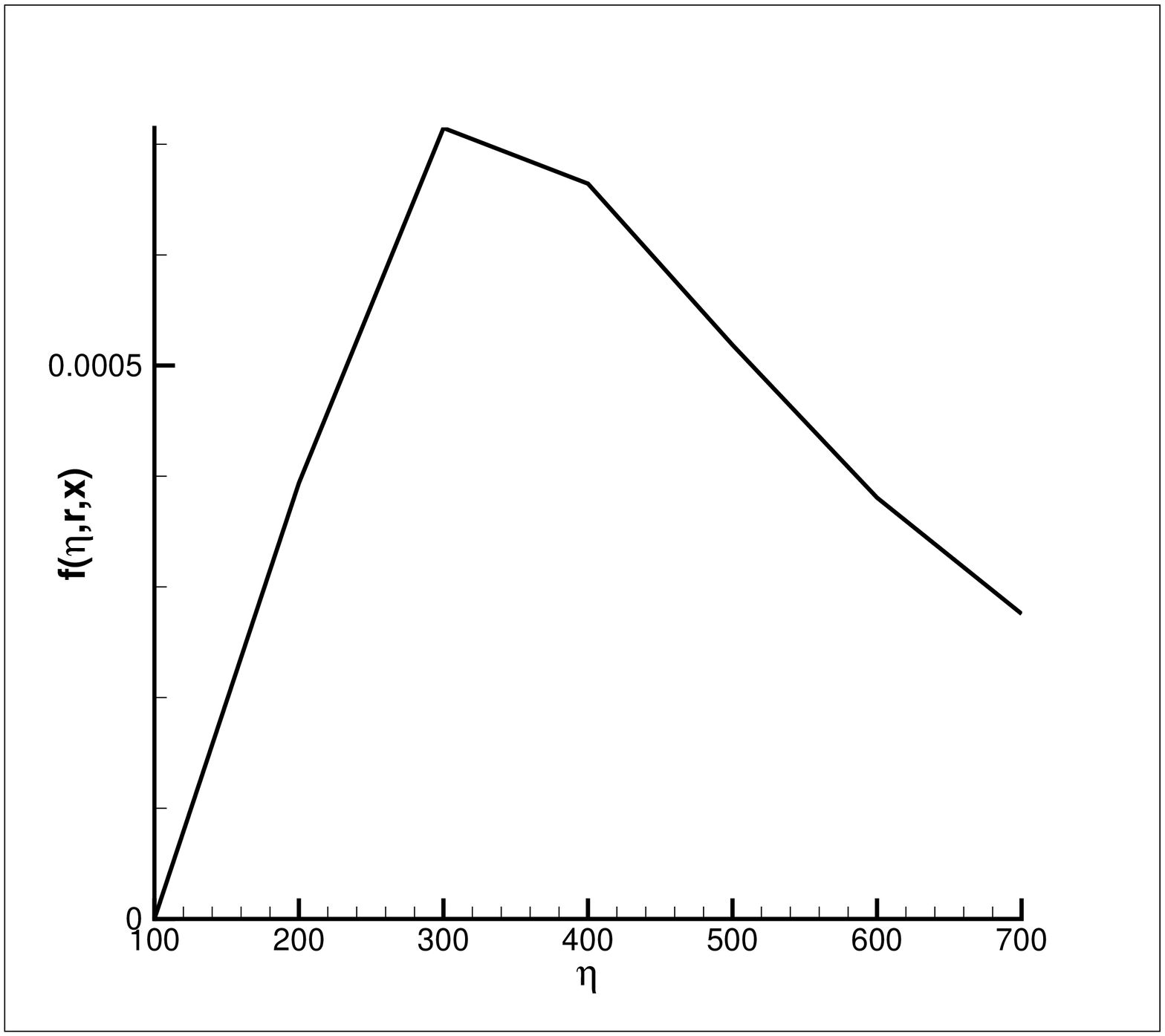}
\end{center}
\begin{center}
\includegraphics[scale=0.21]{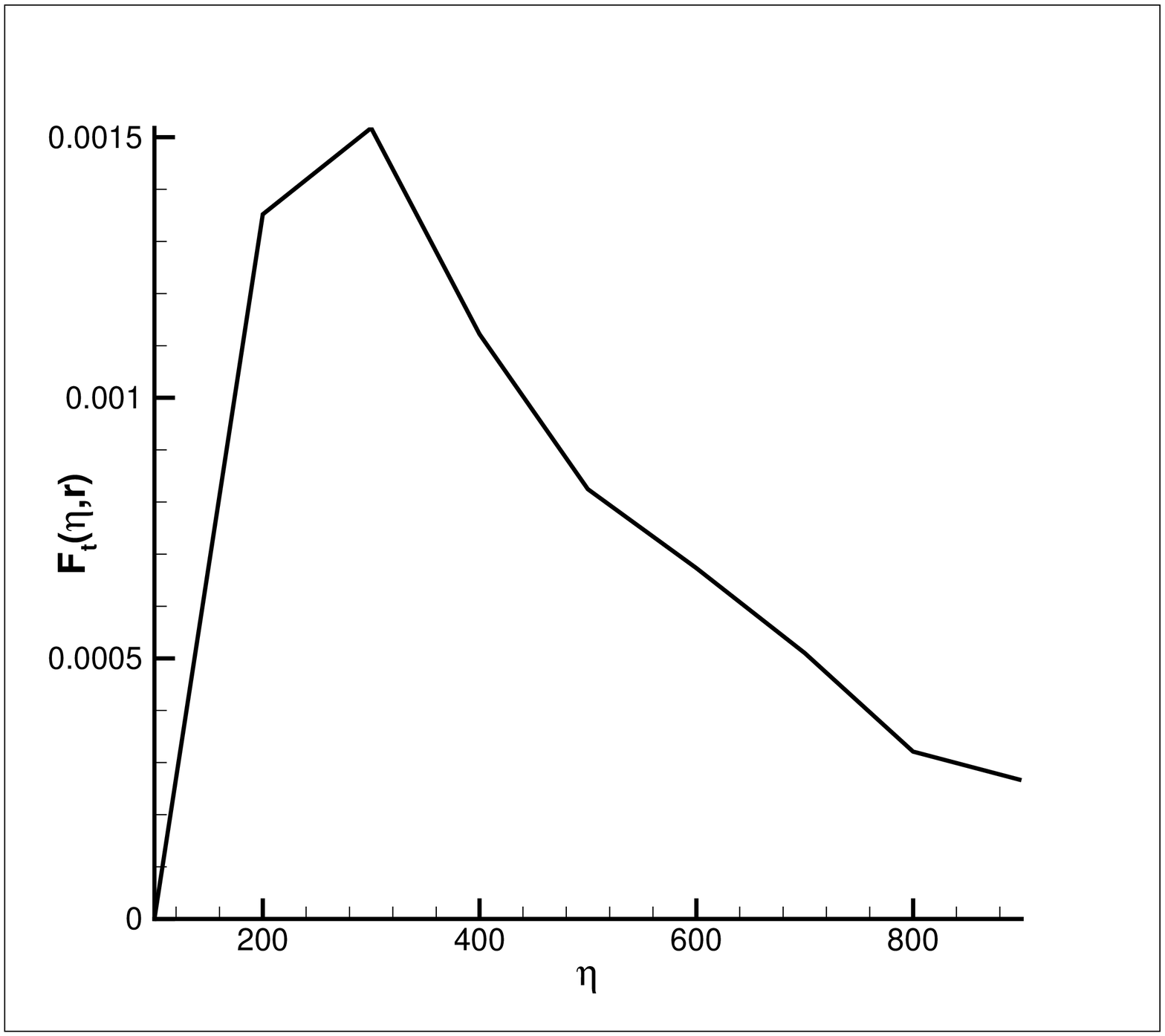}
\includegraphics[scale=0.21]{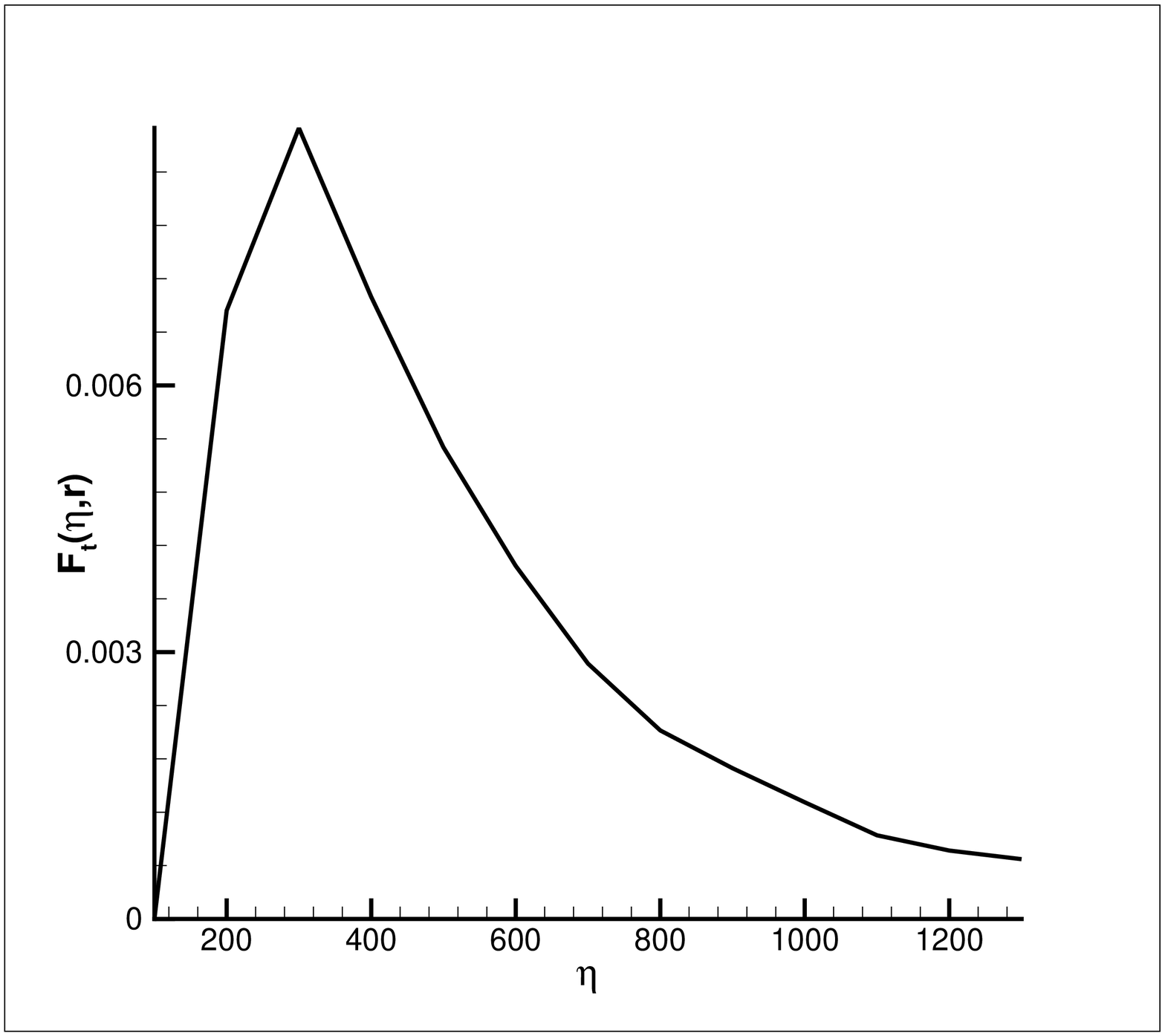}
\includegraphics[scale=0.21]{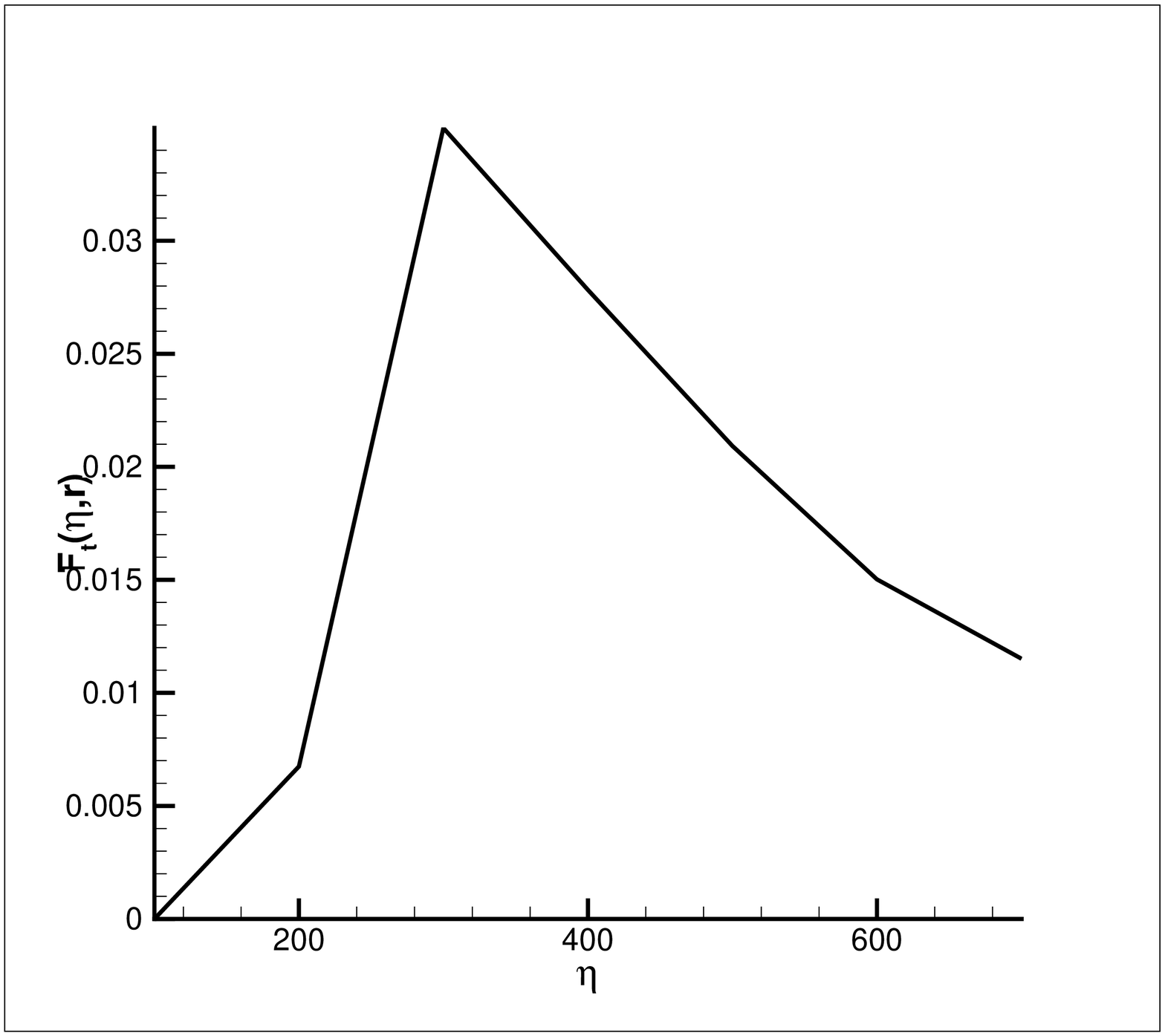}
\caption{Top panel: the time-dependence of $f(\eta, r, x)$ at $x=0$
and at $r=10^2$ of flash source with $\eta_0=1$ (left), 5 (middle) and
$20$ (right). Bottom panel: the light curves of the total flux $F_t(\eta,r)$
of the corresponded top panel.}
\end{center}
\end{figure}

Figure 9 presents the light curves of flash sources with time
duration $\Delta \eta=\eta_0=1$, 5 and 20. The halo size is
$r=10^2$. For the case of $\eta_0=1$, we have $\eta_0\ll r$.
Therefore, the source can be considered as a pulse. The top panels
of Figure 9 are for flux of $\nu_0$ photons, while the bottom panels
are the corresponded total flux. Although the three flash sources
have very different time durations at $r=0$, their light curves of
$f$ at $x=0$ are very similar. The maximum values of $f$ for
$\eta_0=1$, 5 and 20 can even be described by relations as
$f_{20}\simeq 4f_{5}\simeq 20 f_{1}$. The coefficients  4 and 20 are
from the ratio of the total numbers of photons of the three flashes.
The three light curves of $F_t$ at the bottom panels of Figure 9 are
also similar from each other, although they are not as good as the
top three curves $f$. This is because the light curves are
frequency-dependent. Nevertheless, we still see the three maximum
values of $F_t$ also roughly satisfy the relation $F_{20}\simeq
4F_{5}\simeq 20 F_{1}$.

Either Figure 8 or Figure 9 reveals that the time scales of the propagation of a
flash in halos are mainly dependent on the size $r$, regardless the original time duration.
This feature indicates that the spatial transfer of a flash
essentially is a diffusion process. As mentioned in \S 3, the
spatial transfer of resonant photons cannot be described as a purely
Brownian diffusion process, by which the time duration $\Delta \eta$
should increase with $r^2$ or $\tau^2$. On the other hand, if the
spatial transfer of a photon can be, in average, described by
constant speed, the time duration $\Delta \eta$ of a flash should
be, in average, equal to the original time duration, or at least, dependent on the
original time duration. However, Figures 8 and 9
shows that the time duration $\Delta \eta$ is roughly proportional
to $r$, and independent of the initial time duration. Therefore, the
spatial transfer of resonant photons essentially should still be
a diffusion process.

\subsection{Delayed emission line of stored photons}

The effect of trapping and storing photons in an optically thick halo
can be more clearly revealed with a flash source. Let us consider a
flash source with a continuant spectrum, which is described by
equation (17), but we take $S_0\phi_s(x)=1$. The time evolutions of $j$
and $f$ for $\eta_0=50$ at $r=10^2$ are presented in Figure 10, in
which we take the time to be $\eta = 200$, 300 and 500.

\begin{figure}[htb]
\begin{center}
\includegraphics[scale=0.21]{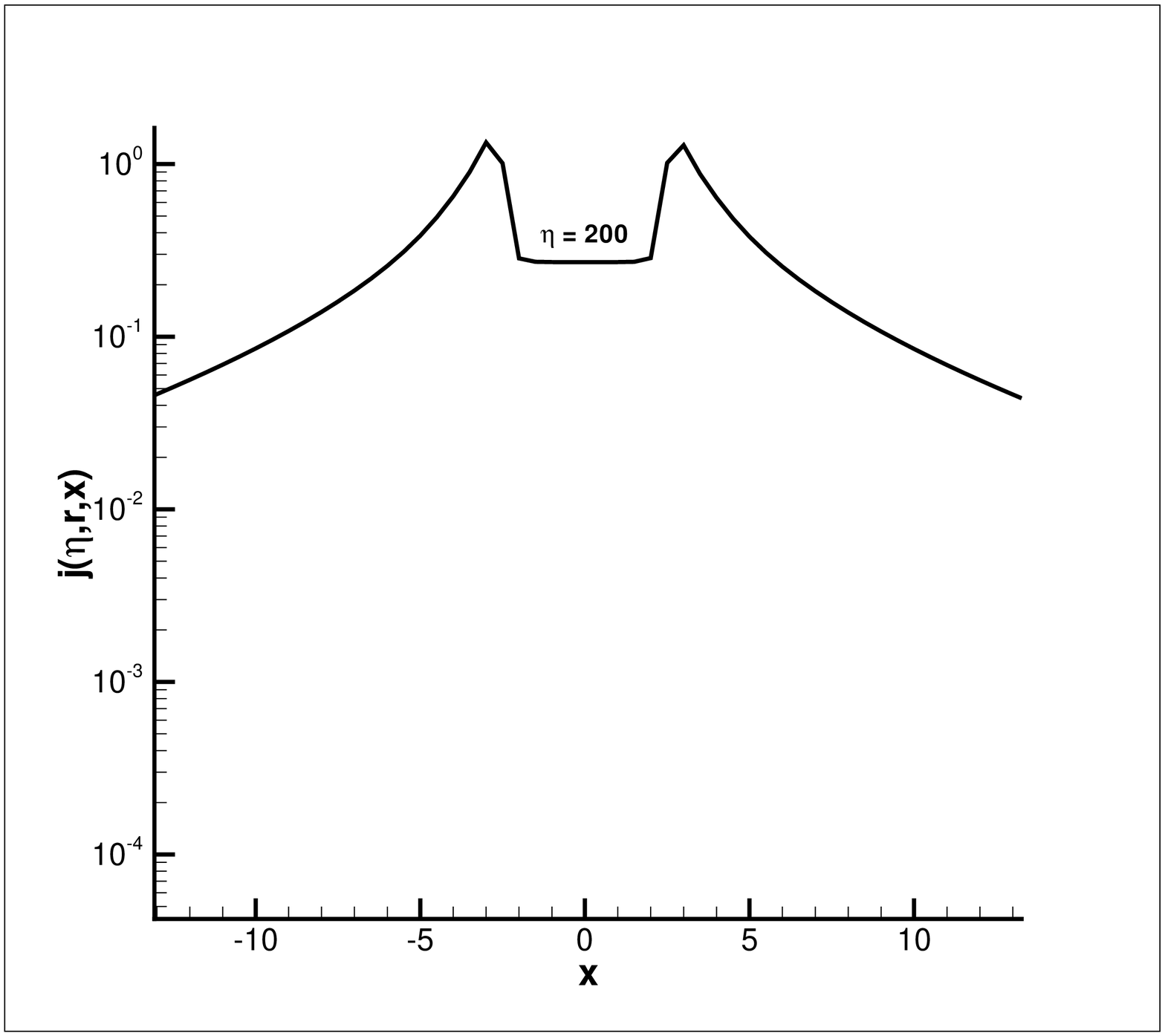}
\includegraphics[scale=0.21]{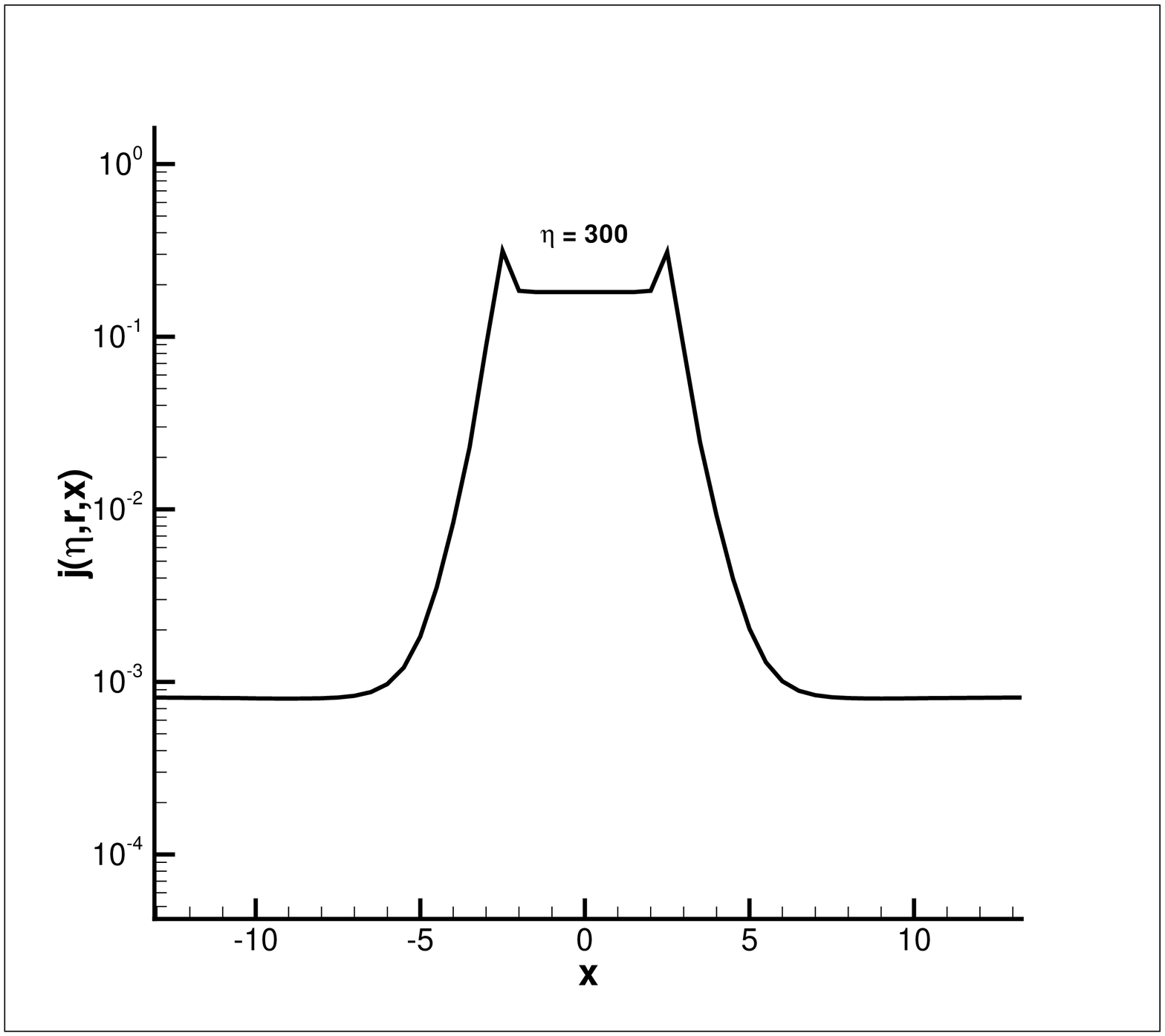}
\includegraphics[scale=0.21]{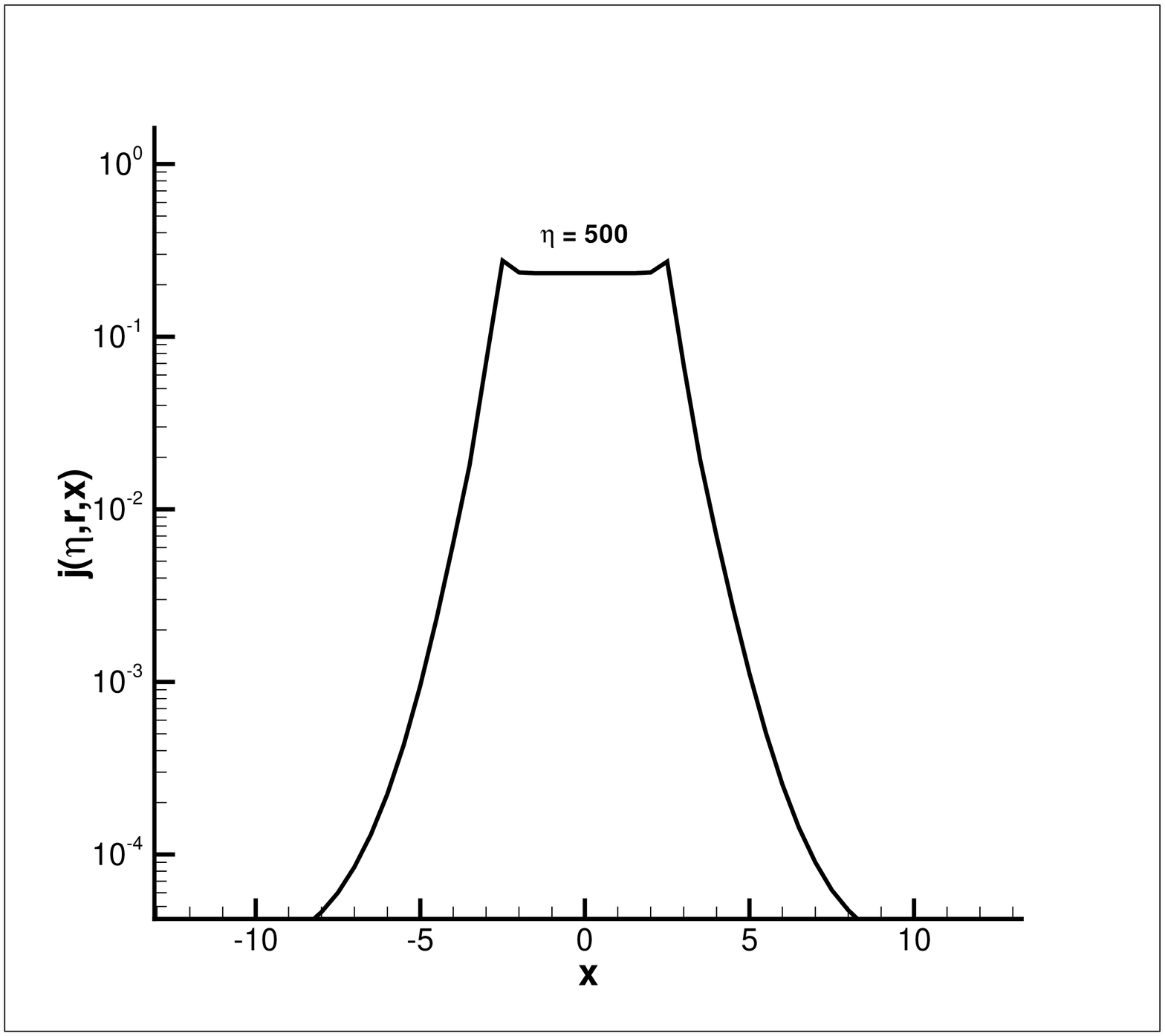}
\end{center}
\begin{center}
\includegraphics[scale=0.21]{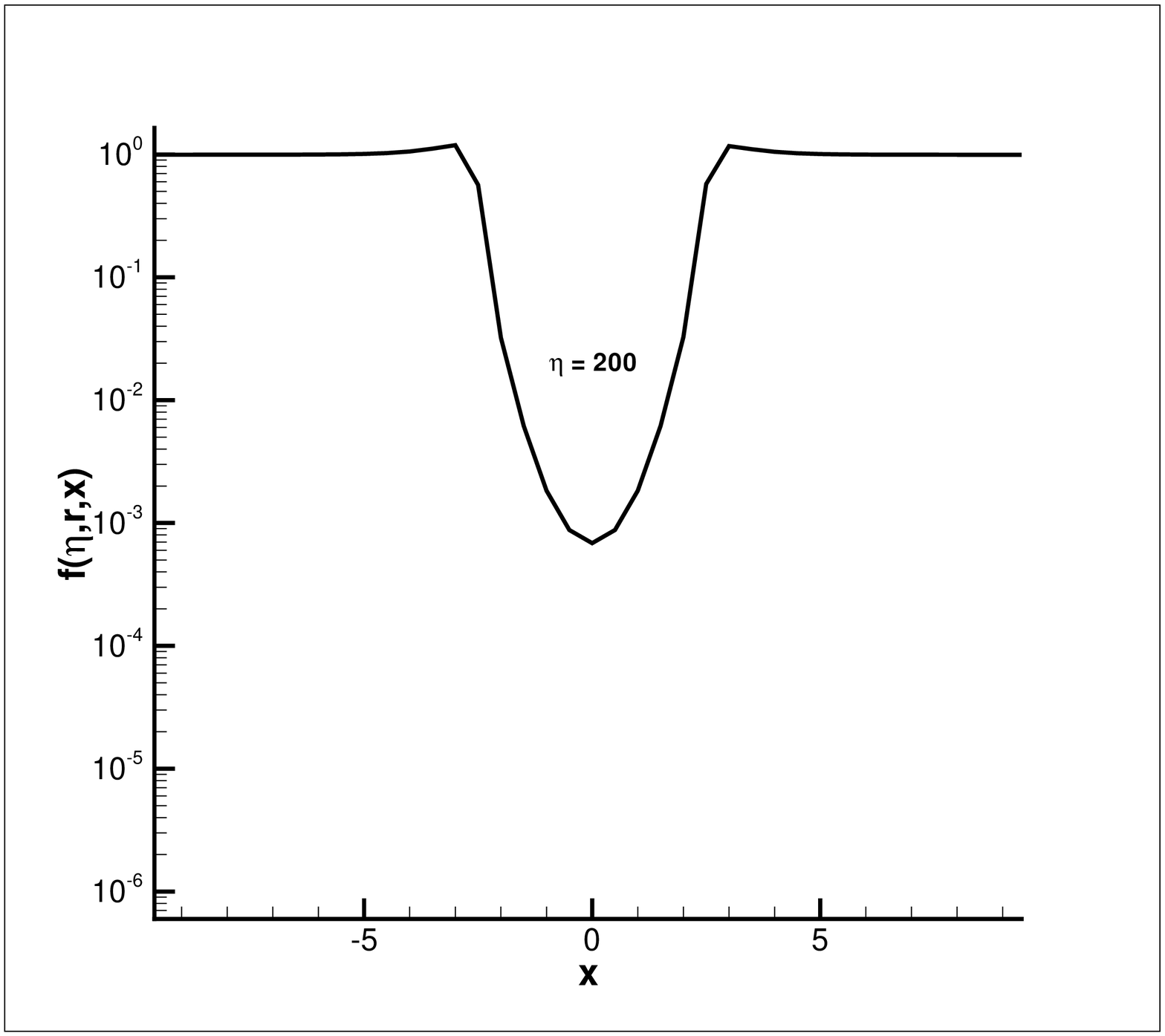}
\includegraphics[scale=0.21]{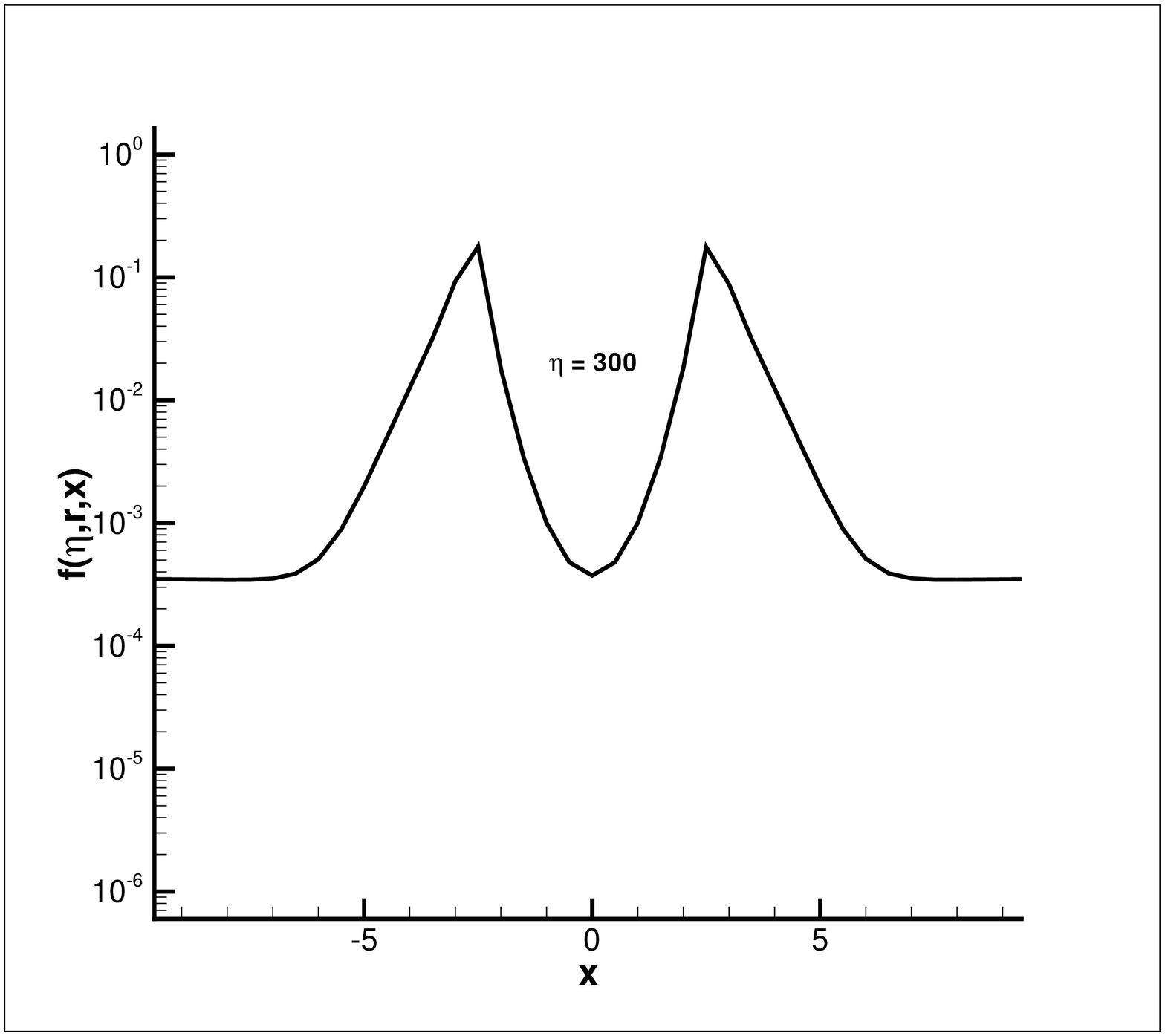}
\includegraphics[scale=0.21]{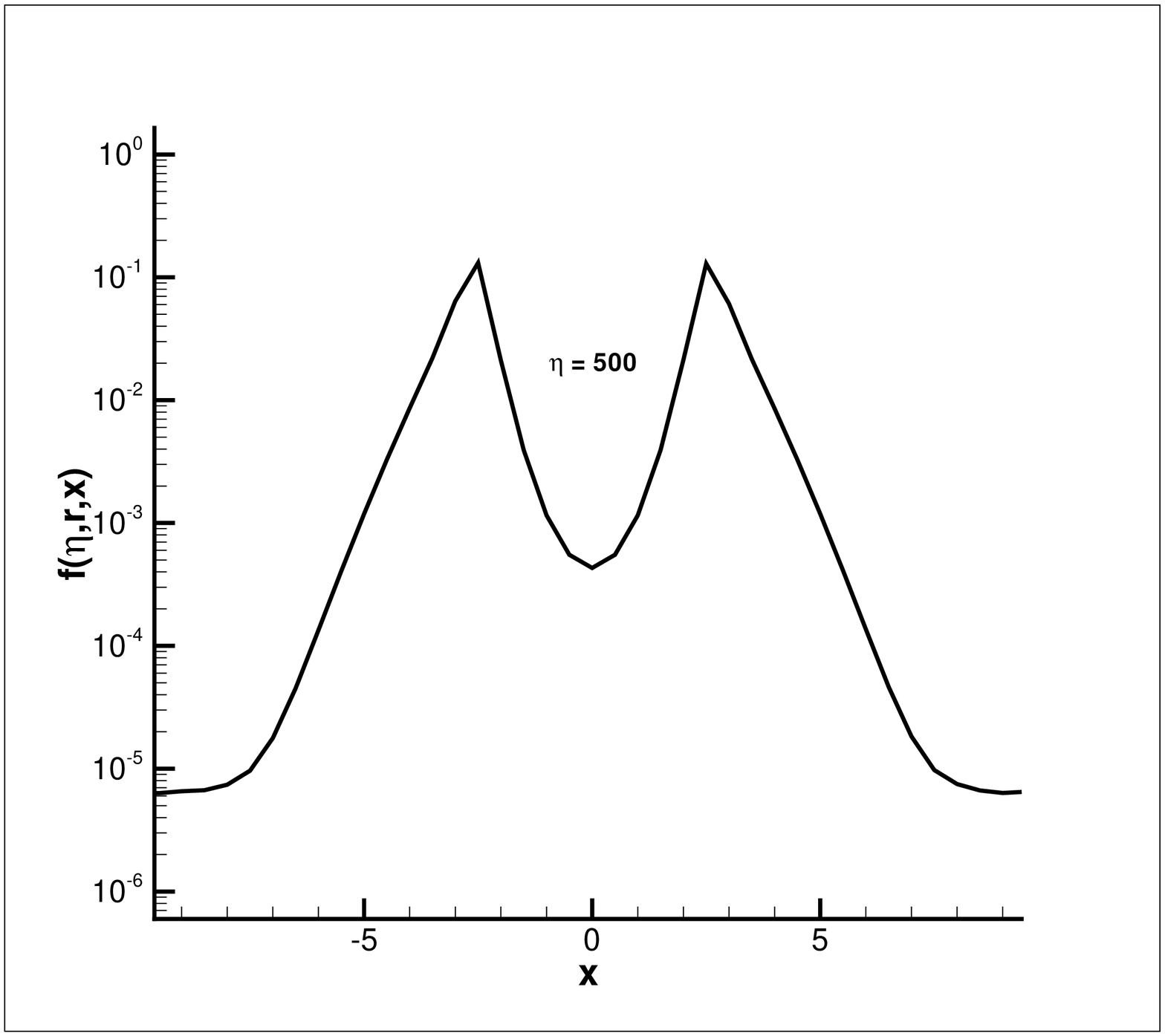}
\end{center}
\caption{Profiles of $j(\eta, r, x)$ (top) and $f(\eta, r,  x)$
(bottom) at $r=10^2$ when a flash source eq.(17) with $\eta_0=50$
and $S_0\phi_s(x)=1$. The time is $\eta=200$ (left), 300 (middle)
and 500 (right). }
\end{figure}

The top panels show the evolution of the mean intensity $j$ in the
halo. In an early time $\eta=200$, there are photons in the central
range $|x|<4$ as well as in wings $|x|>4$. At later time $\eta=300$,
wing photons disappear, because all wing photons from the flash
source have already escaped from the $r$ halo. At time $\eta=500$, the
mean intensity $j$ of $|x|<4$ is still about the same as $j$ at
$\eta=300$. The flat plateau of $j$ is shown in all time. Therefore, the
locally thermalized photons are stored in the halo at least from time
$\eta=200$ to 500, which is much longer than the $\eta_0=50$.

The evolution of the flux $f$ given by the bottom panels of Figure 10 is
more interesting. At the time $\eta=200$, the flux shows a typical
absorption line at $\nu_0$. However, at $\eta=300$, the flux $f$
becomes a typical emission line with two peak profile. At time
$\eta=500$, $f$ is still a two peak emission line. The flux of the
emission at $\eta=500$ is as strong as that at $\eta=300$. Its light
curve is similar to that of Figures 8 and 9. Note that Figures 8 and
9 are for a source of emission line, while Figure 10 is for a
continuant spectrum. The similarity of the light curves of Figure 10
with Figures 8 and 9 is again due to the local thermalization and
the diffusion in the physical space, both of which lead to the initial
frequency spectrum and the time dependence of the photon sources
being forgotten.

The emission at $\eta>300$ is a delayed emission, as the flash of
source has already ceased.  The photons of the delayed emission is
provided by the $|x|<4$ photons stored in the halo $r<10^2$. The
time duration of the delayed emission is about the same as the time
scale of the decaying phase of Figures 8 and 9. Therefore, it is
also proportional to $r$, regardless the original time duration.
Thus, at large $r$, a flash with a continuant spectrum and very
short original time duration $\eta_0$ can produce a two peak
emission with time duration proportional to $r$.

\section{Discussions and conclusions}

The resonant scattering made the transfers of resonant photons in
physical space and frequency space to be coupled from each other. It
leads to the time scale $\eta$ of the spatial transfer of the
resonant photons in the halo with optical depth $\tau \gg 1$ being
much faster than a purely Brownian diffusion process requiring $\eta
\propto \tau^2$. However, essentially it is still a diffusion
process, which can be approximately described by $\eta \propto
\tau^{1/H}$, with the index $H$ less than but very close to 1. It is
possible, if we consider the single longest excursion playing the
role of a long-range dependence, that this diffusion process has a
positive correlation, or is a fractal Brownian diffusion (e.g. Beran
1994).

The number of photons basically is conserved.  Thus, an optically
thick halo is a store of photons with frequency $\sim \nu_0$. The
time scale of the store is the same as the time scale of
the above-mentioned diffusion, i.e. approximately proportional to the
optical depth $\tau$. Moreover, the stored photons are always in the
state of local Boltzmann distribution, even when the mean intensity
is highly time-dependent. The initial conditions are forgotten in
the process of approaching the locally thermal equilibrium. The local
Boltzmann distribution is independent of the frequency spectrum and
time-dependence of the source.

All these features show that the $\nu \simeq \nu_0$ photons play the
central role of radiation transfer with resonant scattering. The
major difference between our solutions and some analytical solutions
(Harrington 1973; Neufeld 1990; Dijkstra et al. 2006) is also around
$x =0$. Since the analytical solutions are based on  the assumption
$\phi(x)=a/\sqrt{\pi}x^2$, and the Gaussian core $e^{-x^2}$ is
ignored, it generally leads to $J(x=0)=0$. With these approximations
the solutions cannot show the effects of restoration and thermalization
of photons around $x=0$.

These basic properties found in our numerical solutions yield the
following features of Ly$\alpha$ photons emergent from an optically
thick halo.

\begin{enumerate}

\item At a given $r$, the profile of the two peaks of the flux is
time-independent, and also independent of the initial profile of the
photons. Therefore, it is impossible to estimate the line profile of
the source from the profile of the flux emergent from an optically
thick halo.

\item The frequencies $|x_{\pm}|$ of the two peaks of the flux is not
less than about 2. This would be useful to estimate the kinetic
temperature of the neutral hydrogen atoms.

\item The resonant scattering makes the flux of the
red damping wing very
different from that of the DLA model. The flux is non-zero at frequency
$\nu_0$ or $x=0$.  These results would be useful to discriminate the
DLA model with models which consider the effect of resonant
scattering.

\item The time scales of light curves of the delayed emission of Ly$\alpha$
photons from a flash source is mainly determined by the optical depth of the
halos. On the other hand, the halo is transparent for high energy photons. A
comparison between the light curves of Ly$\alpha$ photons and high
energy photons would be useful to detect the halo.

\end{enumerate}

For halos with large optical depth, the parameter $\gamma$ is small
even at high redshift. When $\gamma \sim 10^{-5}$, the effect of
cosmic expansion on photon evolution in the frequency space actually
is negligible. All solutions $f(\eta, r, x)$ given in this paper are
almost independent of the Hubble expansion. The effect of $\gamma$
would be important if the considered time scale $\eta$ of $f$ is
comparable with that of the Hubble expansion.

\acknowledgments

This work is supported in part by NSF grant
AST-0506734 and ARO grant W911NF-08-1-0520.

\appendix

\section{Numerical algorithm}

To solve equations (9) and (10) as a system, our computational
domain is $(r, x) \in [0,r_{max}]\times[x_{left},x_{right}]$, where $r_{max}$,
$x_{left}$ and $x_{right}$ are chosen such that the solution vanishes to zero
outside the boundaries.  We choose mesh sizes with grid refinement tests to
ensure proper numerical resolution.
In the following, we describe numerical techniques
involved in our algorithm, including approximations to the spatial derivatives,
integrals in the frequency domain, numerical boundary condition and time evolution.

\subsection{The WENO algorithm: approximations to the spatial derivatives}

The spatial derivative terms in equations (9) and (10) are approximated
by a fifth order finite difference WENO scheme.

We first give the WENO reconstruction procedure in approximating
$\frac{\partial j}{\partial x}$,
\begin{equation}
\frac{\partial j(\eta^n,r_i,x_j)}{\partial x} \approx \frac{1}{\Delta x}
(\hat{h}_{j+1/2}-\hat{h}_{j-1/2}),
\end{equation}
with fixed $\eta = \eta^n$ and $r = r_i$.
The numerical flux $\hat{h}_{j+1/2}$ is obtained by the fifth order
WENO approximation in an upwind fashion, because the wind direction is fixed.
Denote
\begin{equation}
h_j = j(\eta^n,r_i,x_j),    \hspace{25mm} j = -2, -1,\cdots,N+3\\
\end{equation}
with fixed $n$ and $i$. The numerical flux from the WENO procedure is
obtained by
\begin{equation}
\hat{h}_{j+1/2}=\omega_1\hat{h}_{j+1/2}^{(1)}+\omega_2\hat{h}_{j+1/2}^{(2)} +
\omega_3\hat{h}_{j+1/2}^{(3)},\\
\end{equation}
where $\hat{h}_{j+1/2}^{(m)}$ are the three third order fluxes on
three different stencils given by
\begin{eqnarray*}
\hat{h}_{j+1/2}^{(1)} &=& -\frac{1}{6}h_{j-1}+\frac{5}{6}h_{j}+\frac{1}{3}h_{j+1},\\
\hat{h}_{j+1/2}^{(2)} &=& \frac{1}{3}h_{j}+\frac{5}{6}h_{j+1}-\frac{1}{6}h_{j+2},\\
\hat{h}_{j+1/2}^{(3)} &=&
\frac{11}{6}h_{j+1}-\frac{7}{6}h_{j+2}+\frac{1}{3}h_{j+3},
\end{eqnarray*}
and the nonlinear weights $\omega_m$ are given by,
\begin{equation}
\omega_m =
\frac{\check{\omega}_m}{\displaystyle\sum_{l=1}^3\check{\omega}_l},
\hspace{5mm}
\check{\omega}_l = \frac{\gamma_l}{(\epsilon+\beta_l)^2},\\
\end{equation}
where $\epsilon$ is a parameter to avoid the denominator to become
zero and is taken as $\epsilon = 10^{-8}$. The linear weights
$\gamma_l$ are given by
\begin{equation}
\gamma_{1} = \frac{3}{10},\hspace{3mm} \gamma_{2} = \frac{3}{5},
\hspace{3mm} \gamma_{3} = \frac{1}{10},
\end{equation}
and the smoothness indicators $\beta_{l}$ are given by,
\begin{eqnarray*}
\beta_1 &=& \frac{13}{12}(h_{j-1}-2h_{j}+h_{j+1})^2 +\frac{1}{4}(h_{j-1}-4h_{j}+3h_{j+1})^2,\\
\beta_2 &=& \frac{13}{12}(h_{j}-2h_{j+1}+h_{j+2})^2 +\frac{1}{4}(h_{j}-h_{j+2})^2,\\
\beta_3 &=& \frac{13}{12}(h_{j+1}-2h_{j+2}+h_{j+3})^2
+\frac{1}{4}(3h_{j+1}-4h_{j+2}+h_{j+3})^2.
\end{eqnarray*}

To approximate the $r$-derivatives in the system of equations (9) and
(10), we need to perform the WENO procedure based on a characteristic
decomposition. We write
the left hand side of equations (9) and (10) as
\begin{equation}
{\bf u}_t + A {\bf u}_r  \\
\end{equation}
where ${\bf u} = (j, f)^T$ and
\[
A =\left( \begin{array}{cc}
0 & 1 \\
\frac{1}{3} & 0\end{array} \right)\]
is a constant matrix.
To perform the characteristic decomposition,
we first compute the eigenvalues, the right eigenvectors, and the left eigenvectors
of $A$ and denote them by,
$\Lambda$, $R$ and $R^{-1}$.
We then project ${\bf u}$ to the local characteristic fields ${\bf v}$ with ${\bf v}
 = R^{-1}{\bf u}$.
Now ${\bf u}_t + A {\bf u}_r$ of the original system is decoupled as two independent
equations as ${\bf v}_t + \Lambda {\bf v}_r$.
We approximate the derivative ${\bf v}_r$ component by component, each with the correct upwind
direction, with the WENO reconstruction procedure
similar to the procedure described above for $\frac{\partial j}{\partial x}$.
In the end, we transform ${\bf v}_r$ back to the physical space by ${\bf u}_r = R {\bf v}_r$.
We refer the readers to Cockburn et al. 1998 for more implementation details.

\subsection{Adaptive mesh procedure for non-uniform grid}

A fifth order conservative finite difference WENO scheme can only be applied to a uniform
grid or a smoothly varying grid. A smooth transformation,
\begin{equation}
\xi = \xi(r) \\
\end{equation}
gives us a uniform grid in a new variable $\xi$. In this case $\xi$ is sufficiently smooth, i.e.,
it has as many derivatives as the order of accuracy of the scheme. Therefore the left
hand side of the (9) and (10) as
\begin{equation}
{\bf u}_t + A {\bf u}_r  \\
\end{equation}
where ${\bf u} = (j, f)^T$ and
\[
A =\left( \begin{array}{cc}
0 & 1 \\
\frac{1}{3} & 0\end{array} \right)\]
is transformed to
\begin{equation}
{\bf u}_t + A \xi_{r}{\bf u}_{\xi}  \\
\end{equation}
and the WENO derivative approximation is now applied to ${\bf u}_{\xi}$.

\end{document}